\documentclass[amsmath,amssymb,12pt]{article}
\usepackage{jheppub}
\pdfoutput=1
\usepackage[utf8]{inputenc}
\usepackage{amsfonts}
\usepackage{graphicx}
\usepackage{slashed}
\usepackage{xcolor}
\usepackage{tensor}
\usepackage{bbm}
\usepackage{subfig}
\usepackage{ulem}

\title{Magneto-thermal transport implies an incoherent Hall conductivity}
\author[1,2]{Andrea Amoretti,}
\author[2]{Daniel K. Brattan,} 
\author[1,2]{Nicodemo Magnoli,}
\author[1,2]{Marcello Scanavino}

\emailAdd{andrea.amoretti@ge.infn.it}
\emailAdd{danny.brattan@gmail.com}
\emailAdd{nicodemo.magnoli@ge.infn.it}
\emailAdd{marcello.scanavino@ge.infn.it}

\affiliation[1]{Dipartimento di Fisica, Universit\`a di Genova,
via Dodecaneso 33, I-16146, Genova, Italy}
\affiliation[2]{I.N.F.N. - Sezione di Genova, via Dodecaneso 33, I-16146, Genova, Italy}

\abstract{ We consider magnetohydrodynamics with an external magnetic field. We find that in general one must allow for a non-zero incoherent Hall conductivity to correctly describe the DC longitudinal and Hall thermal conductivities beyond order zero in the magnetic field expansion. We apply our result to the dyonic black hole, determining the incoherent Hall conductivity in that case, and additionally prove that the existence of this transport coefficient leads to a significantly better match between the hydrodynamic and AC thermo-electric correlators.}

\begin{document}

\maketitle

\section{Introduction}

{\ Magnetohydrodynamics is a collective theory of hydrodynamic modes coupled to electromagnetic degrees of freedom. It is an effective field theory which describes the long-range correlations of near-equilibrium systems, when the microscopic theory is coupled to a  $U(1)$ gauge field. The electromagnetic field can be dynamical, where the evolution of the gauge field is governed by the Maxwell equations from a given initial configuration, or external where the profile is arbitrary up to satisfying the Bianchi identity. We are interested in the latter.}

{\ In recent times magnetohydrodynamics has been intensively studied. New breakthroughs in the theoretical study of magnetohydrodynamics include, among others things, understanding the deeper underlying symmetries and structures that constrain the transport coefficients and subsequently formulating classification schemes \cite{Haehl:2014zda,Hernandez:2017mch}. There have also been applications to the generalized global symmetry reformulation of hydrodynamics \cite{Grozdanov:2016tdf,Armas:2018zbe,Armas:2018atq,Benenowski:2019ule}. At a more practical level the formalism has been used to analyze the physics of relativistic plasmas \cite{goedbloed_poedts_2004}, as well as to understand the behavior of strongly coupled condensed matter systems \cite{Hartnoll:2007ih,Blake:2014yla,Lucas:2015pxa,Patel:2017mjv,Delacretaz:2019wzh,Amoretti:2019buu}.}

{\ In the earliest formulations of $(2+1)$-dimensional relativistic magnetohydrodynamics \cite{Hartnoll:2007ai,Hartnoll:2007ih,Hartnoll:2007ip,Blake:2014yla,Amoretti:2015gna,Lucas:2015pxa,Blake:2015ina,Blake:2015epa} the entire suite of physically relevant conductivities, electric, thermo-electric and thermal, were given in terms of a single incoherent longitudinal conductivity $\sigma_{0}$ for ``not too strong magnetic fields''. The latter requirement is a consequence of matching holographic and hydrodynamic results. In particular, it was discovered that if one assumes the constitutive relation of \cite{Hartnoll:2007ip,Hartnoll:2007ih} for the charge current, which depends on only a single transport coefficient $\sigma_{\mathrm{0}}$, then one matches precisely the DC electric and thermo-electric conductivities. However, the holographic DC thermal conductivities match the hydrodynamic prediction only in the extreme region where charge density completely suppresses the effect of the magnetic field.}

 We claim that a more appropriate hydrodynamic theory contains two non-trivial charge transport coefficients - the usual $\sigma_{0}$ and an incoherent Hall conductivity $\tilde{\sigma}_{\mathrm{H}}$. To fix these quantities we just use the fact that the diffeomorphism and $U(1)$ gauge Ward identities constrain the small frequency expansion of the charge conductivity \cite{Hartnoll:2007ip,Herzog_2009}. In particular, we note that the $\mathcal{O}(\omega^2)$ piece of the charge correlator relies on the values of the DC thermal conductivities. Thusly, by matching our hydrodynamic correlators at small frequency up to and including $\mathcal{O}(\omega^2)$, we find that $\sigma_{0}$ and $\tilde{\sigma}_{\mathrm{H}}$ can be expressed entirely in terms of two system dependent quantities - the longitudinal ($\kappa_{\mathrm{L}}$) and Hall ($\kappa_{\mathrm{H}}$) thermal DC conductivities -  and the thermodynamics. Consequently our hydrodynamic correlators, being dependent only on $\sigma_{0}$, $\tilde{\sigma}_{\mathrm{H}}$ and the thermodynamics at order one in hydrodynamic derivatives, are also expressed entirely in terms of the same variables. It is important to note that the resultant relations are valid at any order in the magnetic field, provided that one knows $\kappa_{\mathrm{L}}$ and $\kappa_{\mathrm{H}}$ exactly.

{\ In \cite{Jensen:2011xb}, where $(2+1)$-dimensional parity violating hydrodynamics is considered up to an including order one in derivatives, an incoherent Hall conductivity is included in the constitutive relation of the $U(1)$ charge current. However, because the authors of that paper consider $B \sim \mathcal{O}(\partial)$ this Hall conductivity is only non-zero to the order worked at if the theory violates spatial parity in the absence of the magnetic field. The incoherent Hall conductivity we will consider is proportional to the magnetic field - which we take to be order zero in derivatives - and exists in a theory that does not violate spatial parity microscopically. It could potentially appear in the formalism of \cite{Jensen:2011xb} at $\mathcal{O}(\partial^2)$, as a new transport coefficient. In principle, our type of Hall conductivity was allowed for in the appendix of \cite{Delacretaz:2019wzh}, but to our knowledge it has never been shown to be non-zero. Here we provide for the first time an expression for $\tilde{\sigma}_{\mathrm{H}}$ (and also $\sigma_0$) in terms of $\kappa_{\mathrm{L}}$, $\kappa_{\mathrm{H}}$ and the thermodynamics. We eventually verify the validity of our results using gauge/gravity duality, analyzing the simple holographic model of the dyonic black hole. In these kinds of holographic models, analytical formulae for the DC thermo-electric transport coefficients in terms of the thermodynamic data are very well known \cite{Hartnoll:2007ai,Hartnoll:2007ip,Herzog_2009,Amoretti:2013nv,Donos:2014cya,Amoretti:2014mma,Amoretti:2014zha,Amoretti:2014kba,Blake:2014yla,Amoretti:2015gna,Blake:2015ina,Donos:2015bxe,Amoretti:2017xto,Amoretti:2017tbk}. Consequently we have been able, using the known result for $\kappa_{\mathrm{L}}$ and $\kappa_{\mathrm{H}}$, to completely determine the incoherent conductivities $\sigma_0$ and $\tilde{\sigma}_{\mathrm{H}}$, and eventually to compare the complete hydrodynamic correlators to the holographic ones.}

{\ This paper is organized into broadly two distinct sections. In section \ref{sec:hydrosec} we consider in general the theory of a relativistic charged fluid in $(2+1)$-dimensions in the presence of an external magnetic field. After reviewing the Ward identities, we show that the incoherent conductivities $\sigma_0$ and $\tilde{\sigma}_{\mathrm{H}}$, and eventually the hydrodynamic AC charge correlator for general frequencies, are completely determined by thermodynamic quantities and the DC longitudinal and Hall thermal conductivities $\kappa_{\mathrm{L}}$ and $\kappa_{\mathrm{H}}$. In section \ref{sec:dyonic} we return to the tried and tested example of the $(3+1)$-dimensional dyonic black hole and apply our formalism, finding that the system is very well described by the hydrodynamics derived in section \ref{sec:hydrosec}. We conclude the paper with a general discussion of the results obtained in section \ref{sec:discussion}.}

\section{Magnetohydrodynamics}
\label{sec:hydrosec}

{\ Consider a $(2+1)$-dimensional system with a conserved, global $U(1)$ current. The (non-) conservation equations of the stress-energy-momentum (SEM) tensor and charge current are
    \begin{eqnarray}
      \label{Eq:SEMconservation}
      \nabla_{\mu} \langle T^{\mu \nu} \rangle &=& F\indices{^{\nu}_{\mu}} \langle J^{\mu} \rangle \; , \\
      \label{Eq:U1conservation}
      \nabla_{\mu} \langle J^{\mu} \rangle &=& 0 \; . 
    \end{eqnarray}
Here $\langle T^{\mu \nu} \rangle$ and $\langle J^{\mu} \rangle$ refer to the ``total currents'' given by variation of the source terms in the action describing our system. In the presence of an electromagnetic field which is $\mathcal{O}(\partial^{0})$ in derivatives the right hand side of \eqref{Eq:SEMconservation} has an explicit source term.}

{\ We assume the existence of a preferred time-like Killing vector field $u^{\mu}$ and $SO(2)$ rotational invariance. We define the following tensor structures
    \begin{eqnarray}
        \Pi_{\mu \nu} = g_{\mu \nu} + u_{\mu} u_{\nu} \; , 
        \qquad \Sigma_{\mu \nu} = \sqrt{-g} \epsilon_{\mu \nu \rho} u^{\rho} \; , \qquad \\
        \Pi_{\mu \nu} \Pi^{\nu \rho} = \Pi\indices{_{\mu}^{\rho}} \; , \qquad
        \Pi_{\mu \nu} \Sigma^{\nu \rho} = \Sigma\indices{_{\mu}^{\rho}} \; , \qquad
        \Sigma_{\mu \nu} \Sigma^{\nu \rho} = - \Pi\indices{_{\mu}^{\rho}} \; , 
    \end{eqnarray}
where $\epsilon_{\mu \nu \rho}$ is the Levi-Civita symbol with $\epsilon_{012}=1$. With respect to these structures we can define a gauge and Lorentz invariant electric $E^{\mu}$ and magnetic field $B$ by decomposing the field strength tensor into
    \begin{eqnarray}
        F_{\mu \nu} &=& u_{\mu} E_{\nu} - u_{\nu} E_{\mu} + B \Sigma_{\mu \nu} \; . 
    \end{eqnarray}
Similarly we can decompose the stress tensor and the electric current in the following manner
    \begin{eqnarray}
        \langle T^{\mu \nu} \rangle &=& \mathcal{E} u^{\mu} u^{\nu} + \left( \mathcal{P}^{\mu} u^{\nu} + \mathcal{P}^{\nu} u^{\mu} \right) + \mathcal{P} \Pi^{\mu \nu} + \mathcal{T}^{\mu \nu} \; , \\
        \langle J^{\mu} \rangle &=& \mathcal{N} u^{\mu} + \mathcal{J}^{\mu} \; , 
    \end{eqnarray}
where all indices not present on $u^{\mu}$ are transverse and we have defined
    \begin{eqnarray}
        \mathcal{E} = u_{\mu} u_{\nu} \langle T^{\mu \nu} \rangle \; , \qquad \mathcal{P} = \frac{1}{2} \Pi_{\mu \nu} \langle T^{\mu \nu} \rangle \; , \qquad \mathcal{N} = - u_{\mu} \langle J^{\mu} \rangle \; , \qquad \\
        \mathcal{P}^{\mu} = - \Pi\indices{^{\mu}_{\rho}} \langle T^{\rho \nu} \rangle u_{\nu} \; , \qquad
        \mathcal{J}^{\mu} = \Pi_{\mu \nu} \langle J^{\nu} \rangle \; , \qquad \\
        \mathcal{T}^{\mu \nu} = \left( \Pi\indices{^{\mu}_{\sigma}} \Pi\indices{^{\nu}_{\rho}}
                                - \frac{1}{2} \Pi^{\mu \nu} \Pi_{\rho \sigma} \right) \langle T^{\rho \sigma} \rangle  \; . \qquad
    \end{eqnarray}
The two index structure $\mathcal{T}^{\mu \nu}$ is symmetric and traceless.}

\subsection{The diffeomorphism and $U(1)$ gauge symmetry Ward identities}

{\ A key role in our derivation will be played by the Ward identities. Essentially, the presence of an order zero in derivative $O(\partial^{0})$ source in the momentum conservation equation will mean that the thermo-electric and thermal conductivities are completely determined by the electric conductivity.}

{\ The Ward identities for the two point functions of the SEM tensor and charge current, on a flat spacetime with a non-zero electromagnetic field \cite{Herzog_2009}, are
    \begin{eqnarray}
        0 &=& - k_{\mu} \langle J^{\alpha} T^{\mu \nu} \rangle
              + i F\indices{_{\mu}^{\nu}} \langle J^{\alpha} J^{\mu} \rangle
              + k^{\nu} \langle J^{\alpha} \rangle - k_{\mu} \eta^{\alpha \nu}
              \langle J^{\mu} \rangle \; , \\
        0 &=& k_{\mu} \left( \langle T^{\alpha \beta} T^{\mu \nu} \rangle
              + \eta^{\alpha \nu} \langle T^{\beta \mu} \rangle
              + \eta^{\beta \nu} \langle T^{\alpha \mu} \rangle 
              - \eta^{\mu \nu} \langle T^{\alpha \beta} \rangle \right) \nonumber \\
          &\;& + i \eta^{\beta \nu} F\indices{_{\mu}^{\alpha}} \langle J^{\mu} \rangle 
               + i \eta^{\alpha \nu} F\indices{_{\mu}^{\beta}} \langle J^{\mu} \rangle
               - i F\indices{_{\mu}^{\nu}} \langle T^{\alpha \beta} J^{\mu} \rangle \; , 
    \end{eqnarray}
where $k^{\mu}=(\omega, \vec{k})$ is the momentum. Contracting with the fluid velocity or spatial projector while specializing to zero wavevector we find that these identities can be written as
    \begin{eqnarray}
              \omega \langle \mathcal{J}^{\mu} \mathcal{P}^{\nu} \rangle
          &=& - \omega  \mathcal{N} \Pi^{\mu \nu} - i E^{\nu} \langle \mathcal{J}^{\mu} \mathcal{N} \rangle
             + i B \Sigma\indices{_{\rho}^{\nu}} \langle \mathcal{J}^{\mu} \mathcal{J}^{\rho} \rangle
             \; , \\
              \omega \langle \mathcal{P}^{\rho} \mathcal{P}^{\sigma} \rangle
          &=& - \left( \omega \mathcal{E} - i E_{\mu} \langle \mathcal{J}^{\mu} \rangle \right) \Pi^{\rho \sigma}
              - i E^{\sigma} \langle \mathcal{P}^{\rho} \mathcal{N} \rangle 
              + i B \Sigma\indices{_{\mu}^{\sigma}} \langle \mathcal{P}^{\rho} \mathcal{J}^{\mu} \rangle \; .
    \end{eqnarray}
In the case of the dyonic black hole that we investigate later, we will take $E^{\mu} \equiv 0$ and $B$ to be constant. Evaluating the Ward identities with these restrictions causes terms proportional to $E^{\mu}$ to drop out. Further, replacing $\mathcal{P}^{\mu}$ by the spatially projected canonical heat current $\mathcal{Q}^{\mu} = \mathcal{P}^{\mu} - \mu \mathcal{J}^{\mu}$ we arrive at the following relations
    \begin{eqnarray}
             \label{Eq:WardOne}
             \langle \mathcal{J}^{\mu} \mathcal{Q}^{\nu} \rangle
          &=& - \mathcal{N} \Pi^{\mu \nu} - \left( \mu \Pi\indices{_{\rho}^{\nu}} - \frac{i B}{\omega} \Sigma\indices{_{\rho}^{\nu}}  \right)
                \langle \mathcal{J}^{\mu} \mathcal{J}^{\rho} \rangle \; , \\
                \label{Eq:WardTwo}
                \langle \mathcal{Q}^{\mu} \mathcal{Q}^{\nu} \rangle
          &=& - \left( \mathcal{E} + \mathcal{P} - \mathcal{N} \mu \right) \Pi^{\mu \nu}
              - \left( \mu \Pi\indices{_{\rho}^{\nu}} - \frac{i B}{\omega} \Sigma\indices{_{\rho}^{\nu}}  \right) \langle  \mathcal{Q}^{\mu} \mathcal{J}^{\rho} \rangle \; . 
    \end{eqnarray}
The importance of the terms which depend on $B/\omega$ cannot be overstated. They are essential to the structure of the correlation functions as they mix different orders in frequency between the correlators. Consequently, knowing the complete AC behavior of the charge conductivity is sufficient to determine the thermo-electric and thermal conductivities.}

{\ More explicitly, we define the AC electric, thermo-electric and thermal conductivities to be
    \begin{eqnarray}
      \label{Eq:ACchargeconductivity}
      \langle \mathcal{J}^{\mu} \mathcal{J}^{\nu} \rangle &=& i \omega \sigma^{\mu \nu}(\omega) \; , \\
      \label{Eq:ACthermochargeconductivity}
      \langle \mathcal{J}^{\mu} \mathcal{Q}^{\nu} \rangle &=& i \omega \alpha^{\mu \nu}(\omega) \; , \\  
      \label{Eq:ACthermoconductivity}
      \langle \mathcal{Q}^{\mu} \mathcal{Q}^{\nu} \rangle &=& i \omega \kappa^{\mu \nu}(\omega) \; ,
    \end{eqnarray}
respectively. In terms of these totally transverse tensor structures, the Ward identities become
    \begin{eqnarray}
              \alpha^{\mu \nu}
          &=& i \frac{\mathcal{N}}{\omega} \Pi^{\mu \nu} - \left( \mu \Pi\indices{_{\rho}^{\nu}} - \frac{i B}{\omega} \Sigma\indices{_{\rho}^{\nu}}  \right) \sigma^{\mu \rho} \; , \\
              \kappa^{\mu \nu}
          &=& \frac{i}{\omega} \left( \mathcal{E} + \mathcal{P} - \mathcal{N} \mu \right) \Pi^{\mu \nu}
              - \left( \mu \Pi\indices{_{\rho}^{\nu}} - \frac{i B}{\omega} \Sigma\indices{_{\rho}^{\nu}}  \right) \alpha^{\mu \rho} \; .
    \end{eqnarray}
As the thermo-electric and thermal conductivities are given entirely in terms of the charge conductivity, and the Ward identities hold for all frequencies, it follows that complete specification of the charge conductivity at all frequencies is sufficient to determine the other two conductivities.}

{\ The microscopic theory has $SO(2)$ rotational invariance with $B$ breaking spatial parity. Consequently we can decompose the conductivity tensor structures into
    \begin{eqnarray}
     \label{Eq:Wardatloww}
     (\sigma(\omega), \alpha(\omega), \kappa(\omega))^{\mu \nu} &=& (\sigma_{\mathrm{L}}, \alpha_{\mathrm{L}}, \kappa_{\mathrm{L}}) \Pi^{\mu \nu}
                                  + \frac{1}{B} (\sigma_{\mathrm{H}}, \alpha_{\mathrm{H}}, \kappa_{\mathrm{H}}) \Sigma^{\mu \nu} \; ,
    \end{eqnarray}
where the tensors $\Pi^{\mu \nu}$ and $\Sigma^{\mu \nu}$ are treated as order zero in fluctuations, namely substituting $u^{\mu}=(1,\vec{0})$. In terms of this decomposition, and at low frequencies, we discover that 
    \begin{eqnarray} 
     \label{Eq:WardExp4sigmaL}
     \sigma_{\mathrm{L}}(\omega) &=& - i \left( \frac{\mathcal{E}+\mathcal{P}}{B^2} \right) \omega + \frac{\kappa_{\mathrm{L}}(0)}{B^2} \omega^2 + \mathcal{O}(\omega^3) \; , \qquad \\
     \label{Eq:WardExp4sigmaH}
     \sigma_{\mathrm{H}}(\omega) &=& \mathcal{N} 
     + \frac{1}{B^2} \left( \kappa_{\mathrm{H}}(0) + \mu \left(  2 (\mathcal{E}+\mathcal{P}) - \mu \mathcal{N} \right) \right) \omega^2 + \mathcal{O}(\omega^3) \; ,
    \end{eqnarray} 
where we have used that the conductivities must be finite at vanishing $\omega$ and we have assumed that $\mathcal{N}$, $\mathcal{E}$ and $\mathcal{P}$ are independent of frequency up to and including $\mathcal{O}(\omega^2)$. In particular, requiring finite behavior as $\omega \rightarrow 0$ in the Ward identities\footnote{The magnetic field gaps excitations of the system to be at or above the cyclotron frequency in energy. Consequently one expects a smooth limit at low frequencies. This should be compared to relativistic charged hydrodynamics without a background field strength where there is a known $\delta$-function singularity at low frequencies.} constrains
    \begin{eqnarray}
     \label{Eq:WardConstantsatloww}
     \sigma_{\mathrm{L}}(0) = \alpha_{\mathrm{L}}(0) = 0 \; , \qquad \sigma_{\mathrm{H}}(0) = \mathcal{N} \;, \qquad \alpha_{\mathrm{H}}(0) = \mathcal{E} + \mathcal{P} - \mu \mathcal{N} \; , \qquad
    \end{eqnarray}
but leaves $\kappa_{\mathrm{L}}(0)$ and $\kappa_{\mathrm{H}}(0)$ unconstrained and system dependent.  In section \ref{sec:dyonic} we will set them to be the values befitting the dyonic black hole.}

{\ We emphasise that we have not made any magnetization subtractions in our definition of the spatially projected currents and therefore the transport coefficients refer to the total current and not the ``free current''. Moreover we have ignored any of the normalizations by temperature often made to the thermal conductivity so as to not clutter notation.}

\subsection{Equilibrium magnetohydrodynamics}

{\ A comprehensive derivation of the equilibrium configurations of polarizable matter is given in \cite{Kovtun_2016}; however we shall only need the results to lowest order in derivatives. The equilibrium charge current in a theory with only a non-vanishing magnetic field in the background (and no electric field) has the form
    \begin{eqnarray}
        \label{Eq:Equilibriumcurrent}
        \langle J^{\mu} \rangle &=& \rho u^{\mu} - \nabla_{\nu} M^{\nu \mu} \; , \\
        M^{\mu \nu} &=& - m \epsilon^{\mu \nu \rho} u_{\rho} \ \; ,
    \end{eqnarray}
to all orders in derivatives where $\rho = \frac{\partial \mathcal{F}}{\partial \mu}$ is the charge density, $m=-\frac{\partial \mathcal{F}}{\partial B}$ is the magnetization and $\mathcal{F}$ is the free energy.  For our purposes the equilibrium configuration of the charge current decomposed with respect to the time-like vector at zeroth order in derivatives is
    \begin{eqnarray}
        \mathcal{N} = \rho \; , \qquad \mathcal{J}^{\mu} = 0  \; , 
    \end{eqnarray}
where we have taken a ground state with no vorticity.}

{\ Turning now to the SEM tensor, we identify the following expressions at order zero in derivatives,
    \begin{eqnarray}
        \mathcal{E} = - P + T s + \mu \rho \; , \qquad \mathcal{P} = P - m B \; , \qquad\mathcal{P}^{\mu} = 0 \; , \qquad \mathcal{T}^{\mu \nu} = 0 \; .
    \end{eqnarray}
In the above $P$ is the pressure, $T$ the temperature and $s$ the entropy density. Again we have assumed the electric field vanishes in the background. Our microscopic theory will be conformal such that the trace of the SEM tensor gives
    \begin{eqnarray}
       \mathcal{E} - 2 \mathcal{P} = \left( \varepsilon - 2 P + 2 m B \right) = 0 \; ,
    \end{eqnarray}
where $\varepsilon$ is the energy density. We note that the equilibrium configuration of the system depends on the external magnetic field $B$; as will the leading terms in the derivative expansion of the transport coefficients. In systems where the magnetic field is extremely weak - such that it can be treated as $\mathcal{O}(\partial)$ in derivatives - the thermodynamic quantities and transport coefficients can still depend on $B$ but this dependence appears as higher order terms in the derivative expansion i.e.~the leading terms in this latter case are $B$ independent.}

\subsection{AC diffusivities in magnetohydrodynamics}

{\ We wish to work to order one in fluctuations about a flat background at constant temperature $T_{b}$, chemical potential $\mu_{b}$ and magnetic field $B$. Let $u^{\mu} = u^{\mu}_{b} + \delta u^{\mu}$, with $u_{b}^{\mu}=(1,\vec{0})$, be the time-like Killing vector field of the system to order one in fluctuations. We require our fluctuation to maintain $u_{\mu} u^{\mu} = -1$; whence it is the case that $\delta u^{\mu}$ needs to be entirely transverse. At this order in fluctuations the conservation equations have the form
    \begin{eqnarray}
      \partial_{\mu} \delta \langle T^{\mu \nu} \rangle &=& F_{b}^{\nu \mu} \delta \langle J_{\mu} \rangle + \delta F^{\mu \nu} \langle J_{\mu}^{b} \rangle \; , \\
      \partial_{\mu} \delta \langle J^{\mu} \rangle &=& 0 \; .
    \end{eqnarray}
Just as for the full currents, the fluctuations can be decomposed with respect to a time-like vector field. In this case it is useful to use $u_{b}^{\mu}$. Consequently, we can identify
    \begin{eqnarray}
       \delta \langle T^{\mu \nu} \rangle &=& \delta \mathcal{E} u_{b}^{\mu} u_{b}^{\nu} + \left( \delta \mathcal{P}^{\mu} u_{b}^{\nu} + \delta \mathcal{P}^{\nu} u_{b}^{\mu} \right) + \delta \mathcal{P} \Pi_{b}^{\mu \nu} + \delta \mathcal{T}^{\mu \nu} \; , \\
       \delta \langle J^{\mu} \rangle &=& \delta \mathcal{N} u_{b}^{\mu} + \delta \mathcal{J}^{\mu} \; , \\
       \delta F^{\mu \nu} &=& u^{\mu}_{b} \delta E^{\nu} - u^{\nu}_{b} \delta E^{\mu} \; . 
    \end{eqnarray}
With these expressions we can decompose the spatial part of the SEM tensor (non-) conservation equation into the following form
    \begin{eqnarray}
          u_{b}^{\mu} \partial_{\mu} \delta \mathcal{P}^{\nu}
      &=& - \Pi_{b}^{\nu \mu} \left( \partial_{\mu} \delta \mathcal{P} + \partial_{\mu} \delta \mathcal{T}\indices{^{\mu\nu}} \right)
          + \mathcal{N}_{b} \delta E^{\nu}
          + B \Sigma^{\nu \mu}_{b} \delta \mathcal{J}_{\mu} \; ,
    \end{eqnarray}
This will be the only relevant differential equation that we need to solve.}

{\ An unusual feature of any hydrodynamic theory with an explicitly sourced momentum term is the ability to work in the diffusive sector assuming vanishing wavevector $\vec{k}$ from the get-go. This is due to the fact that the diffusive pole does not move to the origin of the complex frequency plane as $\vec{k} \rightarrow 0$. To compare this with ungapped hydrodynamics, the diffusive pole has the form $\omega = - i D \vec{k}^2$ and taking $\vec{k}^2 \rightarrow 0$ in the conservation equations (if one is not careful) gives a trivial result. This inspires us to ignore spatial derivatives in our conservation equations such that the relevant momentum flow equations become
    \begin{eqnarray}
          \label{Eq:MomentumFlow}
          u_{b}^{\mu} \partial_{\mu} \delta \mathcal{P}^{\nu}
      &=& \mathcal{N}_{b} \delta E^{\nu}
          + B \Sigma^{\nu \mu}_{b} \delta \mathcal{J}_{\mu} \; , 
    \end{eqnarray}
for arbitrary - slowly varying - time dependent profiles.}

{\ At the level of linear response we need only determine the fluctuating part of the constitutive relations that are non-zero for completely time dependent profiles. We remind the reader that the electric field is external and permitted to have any time dependence we choose on the condition that the time dependence is sufficiently slow. As such, we will choose it to be a plane wave at a single frequency. With this in mind the constitutive relation for the current takes the form,
    \begin{eqnarray}
      \label{Eq:Currentlinearresponse}
      \delta \mathcal{J}^{\mu}(\omega) &=& \hat{\sigma}_{0}^{\mu \nu} \delta E_{\nu} + \hat{\chi}^{\mu \nu} \delta \mathcal{P}_{\nu}(\omega) \; ,
    \end{eqnarray}
where the subscript $_{0}$ indicates the fundamental (incoherent) conductivity of the theory and the tensor transport coefficients are constant. We have chosen spatial momentum rather than spatial velocity to be one of our fluid variables as it is more convenient for solving the resultant hydrodynamic equations of motion. Spatial rotational invariance allows us to break the transverse tensor structures of \eqref{Eq:Currentlinearresponse} into a piece proportional to $\Pi^{\mu \nu}$ and one proportional to $\Sigma^{\mu \nu}$.}

{\ We would like to highlight a point that will return later, the constitutive relation of \eqref{Eq:Currentlinearresponse} represents the complete response of the charge current in hydrodynamics. We are working at $\vec{k}=0$ so there are no derivative corrections proportional to $\vec{k}$. Moreover one cannot add derivative corrections in $\omega$ without introducing additional modes and taking us outside the hydrodynamic regime. As we are working to order one in fluctuations and both the electric field and momentum vanish in the background there cannot be non-linear tensor structures that correct \eqref{Eq:Currentlinearresponse}. Thus \eqref{Eq:Currentlinearresponse} contains everything consistent with hydrodynamics.}

{\ Applying the definitions of \eqref{Eq:Currentlinearresponse} to \eqref{Eq:MomentumFlow} we see that the spatial momentum (non-) conservation equation becomes
    \begin{eqnarray}
          \label{Eq:MomentumConservation}
          u_{b}^{\mu} \partial_{\mu} \delta \mathcal{P}^{\nu}(t)
      &=& - \Gamma^{\nu \mu} \delta \mathcal{P}_{\mu}(t) + \Theta^{\mu \nu} \delta E_{\nu}(t) \; , 
    \end{eqnarray}
where we have defined
    \begin{eqnarray}
          \Gamma\indices{^{\mu \nu}}
      &=& - B \Sigma\indices{_{b}^{\mu}_{\rho}} \hat{\chi}^{\rho \nu} \; , \\
          \Theta\indices{^{\mu \nu}}
      &=& \mathcal{N}_{b} \Pi_{b}^{\mu \nu} +  B \Sigma\indices{_{b}^{\mu}_{\rho}} \hat{\sigma}_{0}^{\rho \nu}
           \; .
    \end{eqnarray}
It is important to note that, unlike in the case of $B=0$ or $B\sim O(\partial^{1})$, the conservation equation \eqref{Eq:MomentumConservation} is a non-trivial and solvable linear differential equation for the spatial momentum because there is a gap in excitations of the system generated by the magnetic field. In the case of $B=0$ or $B \sim O(\partial^{1})$ the expression on the right hand side of \eqref{Eq:MomentumConservation} vanishes at lowest order in derivatives.}

{\ In the Martin-Kadanoff procedure \cite{KADANOFF1963419} we assume that we turn on some source for our conserved quantities at $t=0$ and allow them to evolve according to the conservation equations. Performing a Laplace transform in time (accounting for boundary conditions at $t=0$) of \eqref{Eq:MomentumConservation} we arrive at
    \begin{eqnarray}
            - i \omega \delta \mathcal{P}^{i}(\omega) - \delta \mathcal{P}^{i}_{0}
        &=& - \Gamma\indices{^{i}_{j}} \delta \mathcal{P}^{j}(\omega)
            + \Theta\indices{^{i}_{j}} \delta E^{j} \; , 
    \end{eqnarray}
where $\delta \mathcal{P}_{0}^{i}$ is the perturbed value of the spatial momentum at $t=0$. Consequently the momentum evolves in frequency according to
    \begin{eqnarray}
            \label{Eq:MomentumEvolution}
            \delta \mathcal{P}(\omega)
        &=& \left( \Gamma - i \omega \mathbbm{1}_{2} \right)^{-1}
            \left( \Theta \delta E + \delta \mathcal{P}_{0} \right) \; ,
    \end{eqnarray}
where indices are implied.}

{\ The frequency evolution of the charge currents can now be determined by substituting \eqref{Eq:MomentumEvolution} into the charge conservation equation employing the constitutive relations \eqref{Eq:Currentlinearresponse}. The result for the charge current is
    \begin{eqnarray}
            \delta \mathcal{J}
        &=&  \left( \hat{\sigma}_{0} + \hat{\chi} \left( \Gamma - i \omega \mathbbm{1}_{2} \right)^{-1} \Theta \right) \delta E +  \hat{\chi} \left( \Gamma - i \omega \mathbbm{1}_{2} \right)^{-1} \delta \mathcal{P}_{0} \; . \qquad
    \end{eqnarray}
From these expressions the frequency evolution of the electric conductivity can be readily determined to be 
    \begin{eqnarray}
         \label{Eq:Completeconductivity}
         \sigma(\omega)
     &=& \hat{\sigma}_{0} + \hat{\chi} \left( \Gamma - i \omega \mathbbm{1}_{2} \right)^{-1} \Theta \; . 
    \end{eqnarray}
To determine the thermal conductivity from the constitutive relations we would in principle need a non-zero spatial momentum. However, we are saved from having to do this by making use of the Ward identities of \eqref{Eq:WardOne} and \eqref{Eq:WardTwo}. }
 
{\ There are some important observations to make about our expressions for the AC diffusivities. Firstly, all poles in these correlation functions must originate in the inverse matrix $\left( \Gamma - i \omega \mathbbm{1}_{2} \right)^{-1}$. The zeroes of the determinant of this matrix will correspond to the quasinormal modes of our dyonic black hole model. Secondly, if we determine the AC response of the charge conductivity, the fundamental conductivities are given entirely in terms of other quantities,
    \begin{eqnarray}
            \label{Eq:TraceRelation1}
            \mathrm{Tr}\left[ \sigma(\omega) \right] 
        &=& \mathrm{Tr}\left[ \hat{\sigma}_{0} \right] - \mathrm{Tr}\left[ \hat{\chi} \left( \Gamma - i \omega \mathbbm{1}_{2} \right)^{-1} \Theta \right] \; , \qquad \\
            \label{Eq:TraceRelation2}
            \mathrm{Tr}\left[ \sigma(\omega) \epsilon \right] 
        &=& \mathrm{Tr}\left[ \hat{\sigma}_{0} \epsilon \right] - \mathrm{Tr}\left[ \hat{\chi} \left( \Gamma - i \omega \mathbbm{1}_{2} \right)^{-1} \Theta \epsilon \right] \; , \qquad \; \;
    \end{eqnarray}
with
    \begin{eqnarray}
            \epsilon
        &=& \left( \begin{array}{cc}
                     0 & 1 \\
                     -1 & 0
                   \end{array} \right) \; , 
    \end{eqnarray}
where we assume we are away from any singularities associated with the inverse operation. In what follows we will decompose our fundamental conductivities as
    \begin{eqnarray}
        \label{Eq:Incoherentdecomposition1}
        \hat{\sigma}_{0}^{ij} &=& \sigma_{0} \delta^{ij} + \tilde{\sigma}_{\mathrm{H}} F^{ij} \; , \\
        \label{Eq:Incoherentdecomposition2}
        \hat{\chi}^{ij} &=& \chi_{0} \delta^{ij} + \chi_{\mathrm{H}} F^{ij} \; , 
    \end{eqnarray}
where we have used spatial parity invariance to argue that the scalar Hall conductivities $\tilde{\sigma}_{\mathrm{H}}$, $\chi_{\mathrm{H}}$ must be  even in $B$ when they multiply the tensor structure $F^{ij}$. We note that unlike some previous formulations \cite{Hartnoll:2007ip,Hartnoll:2007ih,Blake:2015hxa} we have allowed for an incoherent Hall conductivity in \eqref{Eq:Incoherentdecomposition1}. Such a term is not forbidden (in particular by transformations under spatial parity as it multiplies $F^{ij}$) and should therefore be included. In fact it turns out to be necessary. It is consistent with the previous results \cite{Jensen:2011xb} where the magnetic field is treated as $\mathcal{O}(\partial)$ because it would only appear at $\mathcal{O}(\partial^2)$.}

\subsection{Constraining hydrodynamic correlators with the Ward identities}
\label{sec:constraining}

{\ We are now ready to compare the electric conductivities derived in \eqref{Eq:TraceRelation1} and \eqref{Eq:TraceRelation2} to the Ward identities \eqref{Eq:WardExp4sigmaL} and \eqref{Eq:WardExp4sigmaH}, expanding them order by order in the frequency $\omega$. Eventually we constrain the unknown transport coefficients $\sigma_{0}$, $\tilde{\sigma}_{\mathrm{H}}$, $\chi_{0}$ and $\chi_{\mathrm{H}}$ of \eqref{Eq:Incoherentdecomposition1} and \eqref{Eq:Incoherentdecomposition2}.}

{\ The order $\mathcal{O}(\omega^{0})$ equations are trivial so we immediately turn to $\mathcal{O}(\omega^1)$. In the small frequency expansion of the AC correlators, the trace relations \eqref{Eq:TraceRelation1} and \eqref{Eq:TraceRelation2} become
    \begin{eqnarray}
            \frac{i}{2} \mathrm{Tr}[\sigma'(0)]
        &=& \frac{\rho \chi_{0} + \left( \sigma_{0} \chi _{\mathrm{H}} + \chi_{0} \tilde{\sigma}_{\mathrm{H}} \right) B^2}{B^2 \left(\chi_{0}^2 + B^2 \chi_{\mathrm{H}}^2 \right)} \; , \\
            -\frac{i}{2} \mathrm{Tr}[\sigma'(0) F]
        &=& \frac{\sigma_{0} \chi_0 - \rho \chi _{\mathrm{H}} + \chi _{\mathrm{H}} \tilde{\sigma}_{\mathrm{H}} B^2 }{B^2 \left(\chi_{0}^2 + B^2 \chi _{\mathrm{H}}^2\right)} \; . 
    \end{eqnarray}
Comparing the previous expressions with \eqref{Eq:WardExp4sigmaL} and \eqref{Eq:WardExp4sigmaH}, at $\mathcal{O}(\omega^1)$ we can constrain
    \begin{eqnarray}
      \chi_{0} = \frac{\rho -B^2 \tilde{\sigma}_{\mathrm{H}}}{s T + \mu \rho - m B} \; , \qquad
      \chi_{\mathrm{H}} = \frac{\sigma_{0}}{s T + \mu \rho - m B} \; ,
    \end{eqnarray}
which agree with the standard result of \eqref{Eq:StandardIdentificationChiSigma} up to the introduction of the magnetization (a known result) and a fundamental Hall conductivity.}

{\ At $\mathcal{O}(\omega^2)$ we can apply the same process, which will yield expressions for $\sigma_{0}$ and $\tilde{\sigma}_{\mathrm{H}}$ in terms of the DC thermal conductivities $\kappa_{\mathrm{L}}(0)$ and $\kappa_{\mathrm{H}}(0)$ and the thermodynamic variables. The resultant expressions are
  \begin{eqnarray}
        \label{Eq:Fundamentalconductivity}
        \Xi \sigma_{0}(0) &=& (s T + \mu  \rho - m B)^2 \kappa_{\mathrm{L}}(0) \;  , \\
         \label{Eq:FundamentalHallconductivity}
        \Xi \tilde{\sigma}_{\mathrm{H}}(0) &=& - \left(m^2 \left(\kappa_{\mathrm{H}}(0)+\mu ^2 \rho +6 \mu  s T\right)
                                               -\rho  \kappa_{\mathrm{L}}(0)^2\right) \nonumber \\
                                   &\;& + \frac{2 m}{B} 
                                          \left(\kappa_{\mathrm{H}}(0) (s T-\mu  \rho )+\mu  s T (\mu  \rho +3 s T)\right) \nonumber \\
                                   &\;& - \frac{1}{B^2} \left(s^2 T^2-\rho  \kappa_{\mathrm{H}}(0)\right) 
                                          \left(\kappa_{\mathrm{H}}(0)+\mu ^2 \rho +2 \mu  s T\right) 
                                                +2 B \mu  m^3 \; , \\
                                                \label{Eq:Xidef}
                                                \Xi &=& B^2 \left(\kappa_{\mathrm{L}}(0)^2+4 \mu ^2 m^2\right)
                -4 B \mu  m \left(\kappa_{\mathrm{H}}(0)+\mu ^2 \rho +2 \mu  s T\right) \nonumber \\
            &\;& +\left(\kappa_{\mathrm{H}}(0)+\mu ^2 \rho +2 \mu  s T\right)^2 \; . 
    \end{eqnarray}
These expressions are valid to all orders in $B$  and we remind the reader that $\kappa_{\mathrm{L}}(0)$ and $\kappa_{\mathrm{H}}(0)$ are the DC thermal conductivities of the total currents - not the free current. Parenthetically, we note that on the condition $\kappa_{\mathrm{L}}(0) \neq 0$ there is a non-zero $\sigma_{0}$. Moreover, the incoherent Hall conductivity $\tilde{\sigma}_{\mathrm{H}}$ can only be zero if the thermal conductivities are related by the constraint \eqref{Eq:FundamentalHallconductivity} (with $\tilde{\sigma}_{\mathrm{H}}=0$). As we will see in section \ref{sec:dyonic} this is not true in general and as such we generically expect $\tilde{\sigma}_{\mathrm{H}}$ to be non-zero in all but a very special subset of systems.}

{\ Our AC charge conductivity correlator at order one in hydrodynamic derivatives takes the form
 \begin{eqnarray}
            \label{Eq:SigmaL}
            \sigma_{\mathrm{L}}(\omega)
        &=& \frac{i \omega  \left(\gamma_{*}^2+i \gamma_{*} \omega +\omega_{*}^2\right) (s T + \mu \rho - m B)}{B^2 \left((\omega - i \gamma_{*} )^2 - \omega_{*}^2\right)}\; , \\
            \label{Eq:SigmaH}
            \frac{\sigma_{\mathrm{H}}(\omega)}{B}
        &=& \frac{\rho}{B} + \frac{\omega ^2 \omega_{*} ( s T +  \mu  \rho - m B)}
            {B^2 \left((\omega - i \gamma_{*} )^2 - \omega_{*}^2\right)}\; , 
    \end{eqnarray}
where
 \begin{eqnarray}
           \label{Eq:PolePosition1s}
           \omega_{*} 
      &=& \frac{B (s T + \mu \rho - m B) \left(-\kappa _{\mathrm{H}}(0)+2 B \mu  m-\mu  (\mu  \rho +2 s T)\right)}{\Xi} \; , \\
           \label{Eq:PolePosition2s}
          \gamma_{*}
      &=& \frac{B^2 \kappa _{\mathrm{L}}(0) (s T +\mu  \rho - m B)}{\Xi} \; ,
          \end{eqnarray}
and $\Xi$ has been defined in \eqref{Eq:Xidef}.  We include in appendix \ref{app:misc} the AC thermo-electric and thermal conductivities. This is one of our key results as it represents an excellent approximation to the charge correlators that yields the correct values for the DC electric, thermo-electric and thermal conductivities. Moreover, it demonstrates that obtaining the correct DC value of the thermal correlator has nothing to do with including higher order derivative terms nor a frame transformation \cite{Blake:2015hxa} - everything is fixed at $\mathcal{O}(\partial)$ in the constitutive relations, once one takes into account the constraints between the incoherent and the thermal DC conductivities \eqref{Eq:FundamentalHallconductivity}-\eqref{Eq:Xidef}, which are dictated by the Ward identities.}

{\ Since it will be useful in what follows, we also introduce the complexified conductivity,  
    \begin{eqnarray}
     \label{Eq:Complexifiedconductivity}
           \sigma_{+}(\omega) &\equiv& \frac{ \sigma_{H}(\omega)}{B}+i\sigma_{L}(\omega) \nonumber \\
                              &=& B \tilde{\sigma}_{\mathrm{H}} + i \sigma_{0} - \frac{(s T + \mu \rho - m B ) ( \omega_{*} - i \gamma_{*})^2}{B^2 (\omega - (\omega_{*} - i \gamma_{*}))} \; . 
    \end{eqnarray}  
The advantage of this complexified representation is that $\sigma_+$ depends in a straightforward way only on a single hydrodynamic pole located at $\omega_{*} - i \gamma_{*}$, as is evident in \eqref{Eq:Complexifiedconductivity}.}
 
{\ Some notes about \eqref{Eq:Complexifiedconductivity} are rather important. Relativistic hydrodynamics is a derivative expansion in time and space describing the lowest lying quasinormal modes (typically one or two such modes with similar imaginary part). In \cite{Delacretaz:2019wzh} for example there are two constant terms sourcing the momentum and one finds two quasinormal modes are necessary to specify the hydrodynamic limit of the AC conductivity. As hydrodynamics does not incorporate other quasinormal modes, in our case, one expects it to at most get the AC correlator correct to $\mathcal{O}(\omega^2)$. One can motivate this from arguing for the general form of $\sigma_{+}(\omega)$, which is
    \begin{eqnarray}
        \label{Eq:GenericSigma+}
        \sigma_{+}(\omega) &=& \frac{\alpha_{1} + \alpha_{2} \omega}{\omega + \alpha_{3}} \; , 
    \end{eqnarray}
where $\alpha_{1}$, $\alpha_{2}$ and $\alpha_{3}$ are complex numbers. We can use the Ward identities to fix $\alpha_{1}$, $\alpha_{2}$ and $\alpha_{3}$ in terms of the three complex DC conductivities - determining \eqref{Eq:GenericSigma+} uniquely.}

{\ There can be no further corrections in hydrodynamics to \eqref{Eq:Complexifiedconductivity}. Any $\omega$ dependent corrections to \eqref{Eq:Complexifiedconductivity} necessarily introduce additional modes and take us outside the regime of hydrodynamics. Similarly, attempting to improve the position of the quasinormal mode necessarily requires that we modify the $\alpha_{3}$ of \eqref{Eq:GenericSigma+} and subsequently no longer match the DC conductivities\footnote{A discussion of how well our hydrodynamic expression matches the lowest quasinormal mode of the dyonic black hole is relegated to appendix \ref{app:misc}.}. There is in fact a good motivation for fixing the DC conductivities in preference to the quasi-normal mode as one can see that errors in the position of the latter are suppressed by the distance of the complex pole from the real frequency axis. Hence \eqref{Eq:Complexifiedconductivity} is the complete hydrodynamic correlator. Any errors between it and the observed AC conductivity cannot be removed within the hydrodynamic regime.}

\section{Revisiting the dyonic black hole}

\label{sec:dyonic}
{\ We will check the results of the previous section using the holographic dyonic black hole. Eventually, we consider the following action
    \begin{eqnarray}
     \label{Eq:Action}
     S &=& \int d^{3+1}x \; \sqrt{-g} \left( R - 6 - \frac{1}{4} F^2 \right) \; , 
    \end{eqnarray}
where $F$ is a $U(1)$ gauge field strength. The bulk spacetime corresponding to a $(2+1)$-dimensional conformal field theory at strong coupling with a non-zero charge density and magnetic field is the asymptotically AdS$_{4}$ dyonic black hole solution to the equations of motion coming from \eqref{Eq:Action}. This black hole has the metric
    \begin{subequations}
     \label{Eq:Backgroundmetric}
     \begin{eqnarray}
     ds^{2} &=& \frac{dz^2}{f(z)} + \frac{\alpha^2}{z^2} \left( - f(z) dt^2 + dx^2 + dy^2 \right) \; , \\
     f(z) &=& 1 + \left( \rho^2 + B^2 \right) \left(\frac{z}{\alpha}\right)^4 - \frac{1}{\alpha} \left( \alpha^4 + \rho^2 + B^2 \right) \left(\frac{z}{\alpha}\right)^3 \; , 
    \end{eqnarray}
    \end{subequations}
with the horizon at  $z=1$, the boundary at $z=0$ and bulk gauge field strength
    \begin{eqnarray}
      \label{Eq:Backgroundgauge}
      F &=& - \mu \mathrm{d}z \wedge \mathrm{d}t + B \mathrm{d}x \wedge \mathrm{d}y \; .
    \end{eqnarray}
The thermodynamics of this black brane is well known, and here we only list the results. The temperature $T$, the entropy density $s$, the charge density $\rho$ and the magnetization density $m$  are expressed in terms of the bulk data $\mu$, $\alpha$ and $B$ as follows:
    \begin{eqnarray}
       T = \frac{(3 \alpha^4 - \mu^2 - B^2)}{4 \pi \alpha^3} \; , \qquad  \rho = \alpha \mu \; , \qquad m = - \frac{B}{\alpha} \; , \qquad
        s = \pi \alpha^2 \; .
    \end{eqnarray}
As the system is conformally invariant it satisfies a scaling Ward identity which relates the pressure $P$ and the energy density $\varepsilon$:
    \begin{eqnarray}
        \varepsilon = 2 \left( P - m B \right) \; , 
        \qquad \varepsilon = \frac{1}{2 \alpha} \left( \alpha^4 + \rho^2 + B^2 \right) \; .
    \end{eqnarray}
Additionally the system is extensive and therefore satisfies a first law with $\varepsilon + P = \mu \rho + s T $.}

{\ We will be interested in finite frequency fluctuations about the background \eqref{Eq:Backgroundmetric} and \eqref{Eq:Backgroundgauge} corresponding to fluctuations of the boundary electric field. This requires that we consider fluctuations of the $tx$ and $ty$ components, $\delta g_{tx}$ and $\delta g_{ty}$, of the metric and $x$ and $y$ components of the gauge field, $\delta a_{x}$ and $\delta a_{y}$. The analysis of these perturbations at first order in small frequency was completed in \cite{Hartnoll:2007ai}. We record them here
    \begin{eqnarray}
           \langle \mathcal{J}^{\mu} \mathcal{J}^{\nu} \rangle
      &=& - i \omega \frac{\rho}{B} \Sigma^{\mu \nu} + \mathcal{O}(\omega^2) \; , \\
           \langle \mathcal{J}^{\mu} \mathcal{Q}^{\nu} \rangle
      &=& - \frac{i \omega}{B} \left( \frac{3}{2} \varepsilon - \mu \rho \right) \Sigma^{\mu \nu} + \mathcal{O}(\omega^2) \; , \\
            \langle \mathcal{Q}^{\mu} \mathcal{Q}^{\nu} \rangle
      &=& i \omega \left( \frac{(s T)^2}{\rho^2 + B^2} \right) \Pi^{\mu \nu} \nonumber \\
      &\;& - i \omega \left( \frac{\rho}{B( \rho^2 + B^2)} \left( (s T)^2 - ( m^2 + \mu^2 ) B^2 \right) \right) \Sigma^{\mu \nu} + \mathcal{O}(\omega^2)  \; . \qquad
    \end{eqnarray}
These were determined analytically and hold for all values of the magnetic field and charge. Comparing with \eqref{Eq:WardConstantsatloww} we see that we can identify
    \begin{eqnarray}
    \label{Eq:curlydionic}
      \mathcal{N} = \rho \; , \qquad \mathcal{E} + \mathcal{P} = \frac{3 \varepsilon}{2} \; ,
    \end{eqnarray}
and additionally we have
    \begin{eqnarray}
    \label{Eq:analyticalkappadyonic}
      \kappa_{\mathrm{L}}(0) = \frac{(s T)^2}{\rho^2 + B^2} \; , \qquad \kappa_{\mathrm{H}}(0) = \frac{\rho}{\rho^2 + B^2} \left( (s T)^2 - ( m^2 + \mu^2 ) B^2 \right)  \; . 
    \end{eqnarray}
The AdS-CFT correspondence gives the total current as a variation of the on-shell action. Consequently our DC conductivities are with reference to the total current, and not the magnetization subtracted versions that sometimes appear in the literature \cite{Hartnoll:2007ih,Hartnoll:2007ip,Amoretti:2015gna,Blake:2015ina,Blake:2015epa} .}

{\ The Ward identities were demonstrated to hold in the holographic case of the dyonic black hole in \cite{Hartnoll:2007ip}. Through them, should we evaluate the charge conductivity at arbitrary frequency, we will be able to determine the thermo-electric and thermal conductivities. This analysis has been done previously and we refer the reader to \cite{Hartnoll:2007ip}. The result is that the independent response of our theory is described by the coupled bulk equations
    \begin{eqnarray}
     \label{Eq:Bulkeom1}
     f(z) \left( - \rho \mathcal{E}_{+}'(z) + B \mathcal{B}_{+}'(z) \right)
     + \omega \left( B \mathcal{E}_{+}'(z) + \rho \mathcal{B}_{+}'(z) \right) &=& 0 \; , \\
     \label{Eq:Bulkeom2}
     \frac{\omega}{4z^2} \left(  \mathcal{E}_{+}'(z) - \frac{\omega}{f(z)} \mathcal{B}_{+}(z) \right)
     + B^2 \mathcal{B}_{+}(z) - \rho B \mathcal{E}_{+}(z) &=& 0 \; , 
    \end{eqnarray}
where
    \begin{eqnarray}
      \mathcal{E}_{+}(z) &=& i \omega \left( \delta a_{x}(z) + i \delta a_{y}(z) \right)
                             + \frac{i B}{z^2} \left( \delta g_{tx}(z) + i \delta g_{t y}(z) \right) \; , \\
      \mathcal{B}_{+}(z) &=& - B f(z) \left( \delta a_{x}'(z) - i \delta a_{y}'(x) \right) \; . 
    \end{eqnarray}
}

{\ The asymptotic expansion of the fields $\mathcal{E}_{+}(z)$ and $ \mathcal{B}_{+}(z)$ yield the boundary electric field and charge currents respectively,
    \begin{eqnarray}
      \lim_{z \rightarrow 0} \mathcal{E}_{+}(z) = E_{x} + i E_{y} \; , \qquad
       -i \lim_{z \rightarrow 0} \mathcal{B}_{+}(z) = J_{x} + i J_{y} \; . 
    \end{eqnarray}
This provides another motivation for us to consider the complex charge conductivity 
    \begin{eqnarray}
       \lim_{z \rightarrow 0} \frac{\mathcal{B}_{+}(z)}{\mathcal{E}_{+}(z)} = \sigma_{+}(\omega) = \sigma_{xy}(\omega) + i \sigma_{xx}(\omega) \; .
    \end{eqnarray}
 Expressed in terms of this complexified conductivity, and having substituted the dyonic black hole results \eqref{Eq:curlydionic} for $\mathcal{N}, \; \mathcal{E}$ and $\mathcal{P}$, the Ward identities \eqref{Eq:WardExp4sigmaL} and \eqref{Eq:WardExp4sigmaH} give:
 \begin{eqnarray}
     \label{Eq:SigmaPlusLowOmegaExpansion}
     \sigma_{+}(\omega) &=& \frac{\rho}{B} + \frac{s T + \mu \rho - m B }{B^2} \omega
                            \nonumber \\
                        &\;& + \left[ \frac{2 \left(\kappa_{\mathrm{H}}(0)+\mu ^2 \rho +2 \mu  s T - 2 \mu m B \right) }{2 B^3} + i \frac{\kappa_{\mathrm{L}}(0) }{B^2} \right] \omega ^2  +O\left(\omega ^3\right)  \; . \qquad
    \end{eqnarray}

\subsection{An incoherent conductivity}

{\ We now prove that the formulae for the incoherent conductivities given in \eqref{Eq:Fundamentalconductivity} and \eqref{Eq:FundamentalHallconductivity} are actually valid in the dyonic black hole. The usual definition of such a quantity in terms of the charge current orthogonal to momentum \cite{Davison:2015taa} will no longer suffice as the magnetic field $B$ mixes the two spatial components of the momentum. Consequently there is no part of the charge current which is orthogonal to the momentum at all points in space. Instead, we return to the original motivation for defining the incoherent conductivity - it is the contribution to the correlator that is independent of coherent dissipative mechanisms. Such mechanisms when relevant to hydrodynamics can be introduced into the formalism by modifying the source term of the momentum equation by shifting $\Gamma^{ij}$ to $\Gamma^{ij} + \Gamma_{\mathrm{coherent}}^{ij}$.}

\begin{figure}[t]%
    \centering
    \subfloat[]{\includegraphics[width=0.45\textwidth]{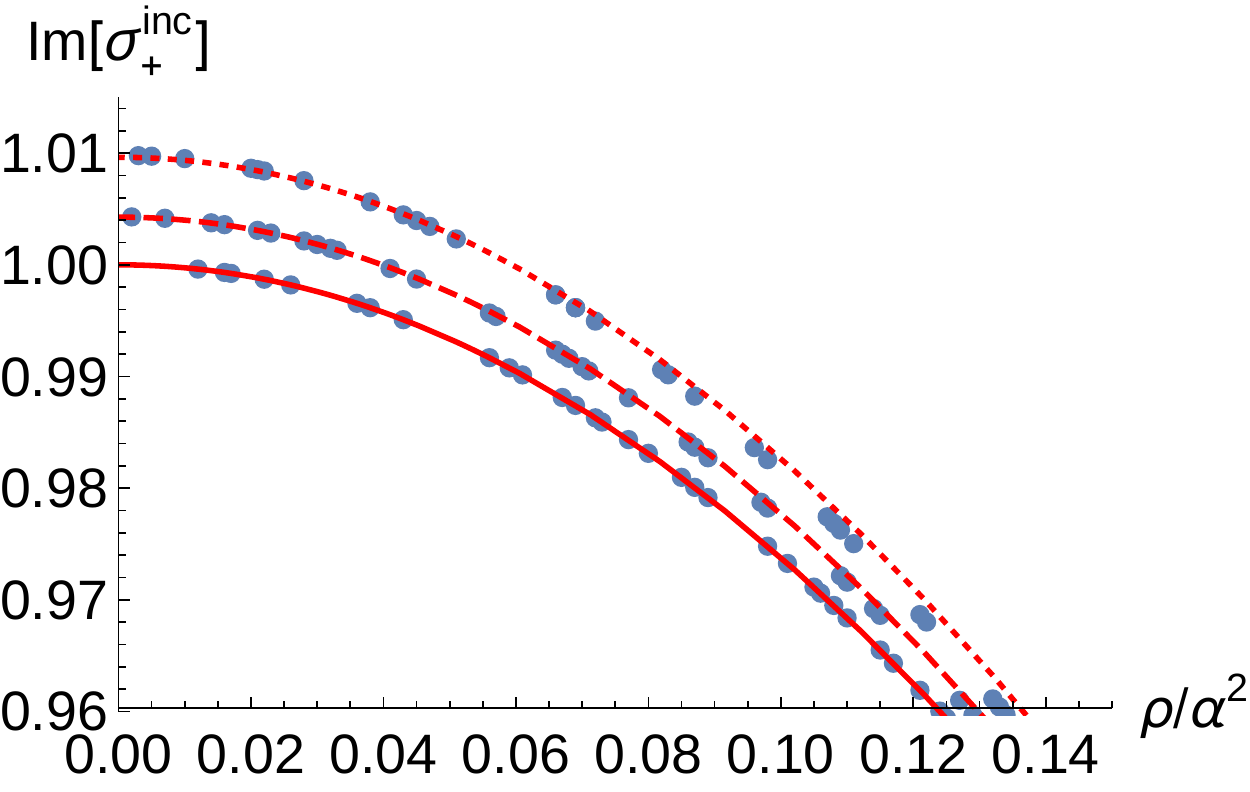}}\qquad
    \subfloat[]{\includegraphics[width=0.45\textwidth]{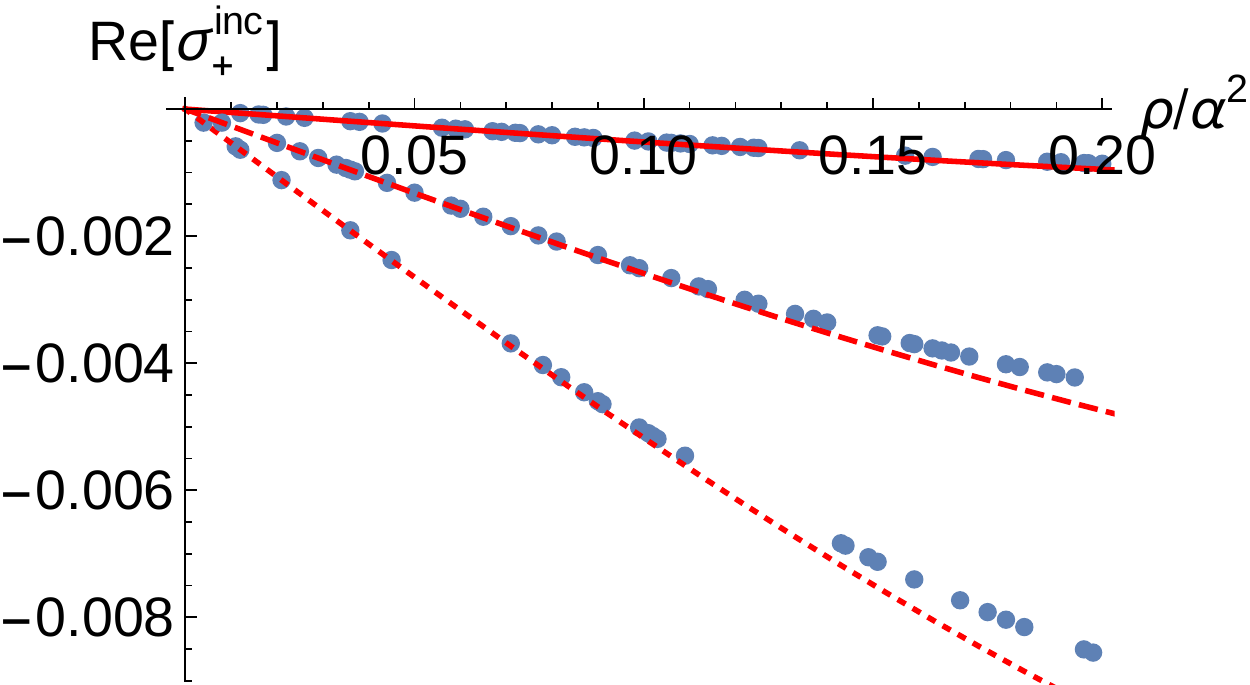}}\\
    \subfloat[]{\includegraphics[width=0.45\textwidth]{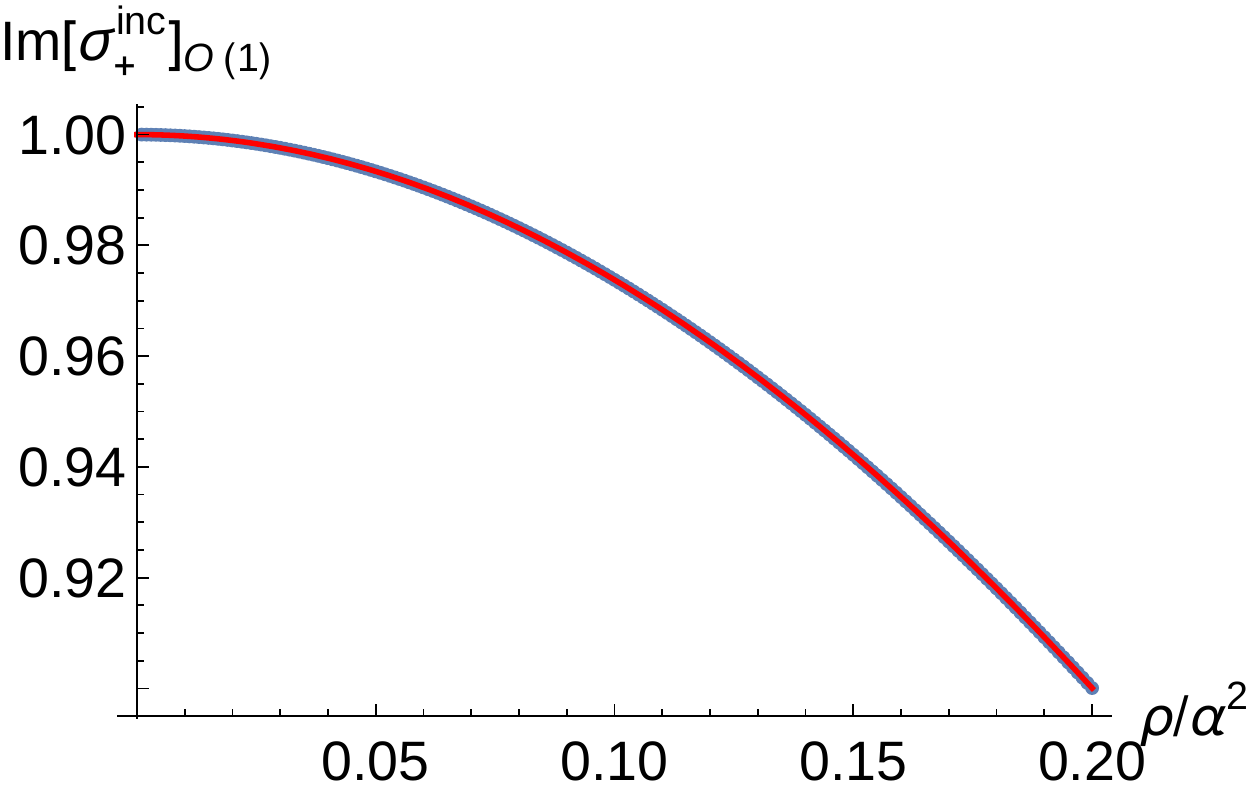}}\qquad
    \subfloat[]{\includegraphics[width=0.45\textwidth]{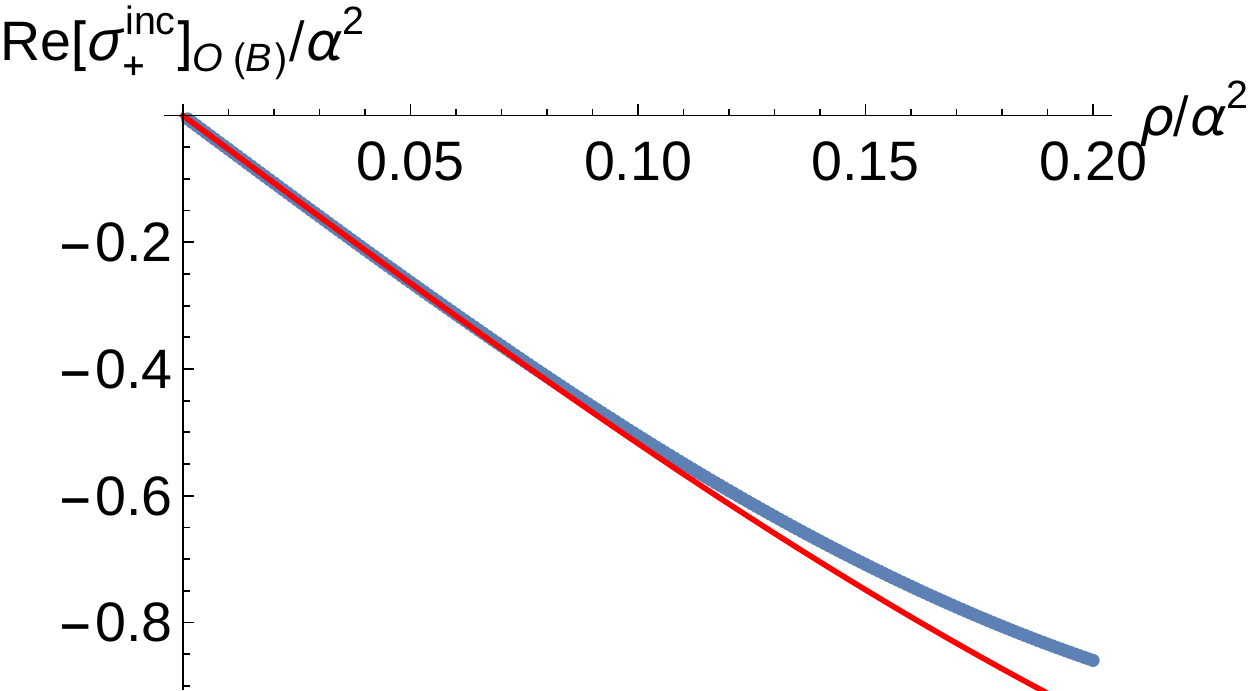}}
    \caption{Plots of the constant term in the Laurent expansion of the charge correlator about the hydrodynamic pole against the charge density. The blue dots are data. \textbf{Upper left:} The imaginary part of $\sigma_{+}^{\mathrm{inc.}}$ against our analytic expression for $\sigma_{0}$. The three red lines represent $B/\alpha^2=1/1000$ (solid), $1/25$ (dashed) and $3/50$ (dotted). Notice that $\sigma_{0}>1$ which stands in contradiction to the standard prescription where $\sigma_{0}=(sT/(\varepsilon +P))^2 \leq 1$. \textbf{Upper right:} The real part of $\sigma_{+}^{\mathrm{inc.}}$ against our analytic expression for $\tilde{\sigma}_{\mathrm{H}}$. The three red lines represent $B/\alpha^2=1/1000$ (solid), $5/1000$ (dashed) and $10/1000$ (dotted). \textbf{Lower left:} The leading contribution at small $B$ to the imaginary part of the constant term. The solid red line is the analytic expression for $[\sigma_{0}]_{B=0}$. \textbf{Lower right:} The $\mathcal{O}(B^1)$ contribution to the constant part of the Laurent expansion. The red line is our analytic result for $[\tilde{\sigma}_{\mathrm{H}}]_{B=0}$.}
    \label{Fig:Incoherentconductivity}
\end{figure}

{\ With this remark in mind we define the incoherent conductivity to be the constant term in the Laurent expansion of the complexified conductivity $\sigma_+$ as defined in \eqref{Eq:Complexifiedconductivity} about the hydrodynamic pole located at $\omega_*-i \gamma_*$ . This is invariant under the translation $\omega \rightarrow \omega - i \Gamma_{\mathrm{coherent}}$ and equal to the first term of \eqref{Eq:Complexifiedconductivity} i.e.
    \begin{eqnarray}
        \label{Eq:Incoherentconductivity}
        \sigma_{+}^{\mathrm{inc.}} \equiv B \tilde{\sigma}_{\mathrm{H}} + i \sigma_{0} \; . 
        \end{eqnarray}}
At lowest order in $B$ these terms are
    \begin{equation}
        \label{Eq:FundamentalconductivitieslowB}
        \left[ \sigma_{0} \right]_{B=0} = \left( \frac{3 \alpha^4 - \rho ^2}{3 (\rho ^2+ \alpha^4 )^2} \right)^2 \; , \; \; 
        \left[ \tilde{\sigma}_{\mathrm{H}} \right]_{B=0} = -\frac{16 \rho  \left(\rho ^2+3 \alpha^4\right) \left(5 \rho ^4+6 \alpha^4 \rho ^2+9 \alpha^8 \right)}{81 \left(\rho ^2+\alpha^4 \right)^4} \; , \qquad
    \end{equation}
when expressed in dyonic black hole data. In fact, it should be noted that $\tilde{\sigma}_{\mathrm{H}}$ vanishes as $\mathcal{O}(\rho)$ independent of the value of $B$ for the dyonic black hole.}

{\ We have checked the validity of the relation \eqref{Eq:Incoherentconductivity} against the numerics. Displayed in the upper plots of fig.~\ref{Fig:Incoherentconductivity} are our analytic expressions for the incoherent conductivities against charge density at various values of the magnetic field. For low magnetic fields the match is excellent as expected, becoming progressively worse as we increase the magnetic field and charge density (and therefore effectively lower the temperature). Moreover, our result for $\sigma_{0}$ becomes greater than one for  $B>\rho$, in agreement with the data. The result from the standard magentohydrodynamic approach to the dyonic black hole leads to an incoherent conductivity $\sigma_{0}$ bounded above by one.}

{\ Additionally, in fig.~\ref{Fig:Incoherentconductivity} we display $\left[ \sigma_{0} \right]_{B=0}$ against the numerically extracted constant Laurent coefficient at low $B$ in the lower left hand plot of fig.~\ref{Fig:Incoherentconductivity} and the matching is excellent. In the lower right plot we also show $\left[ \tilde{\sigma}_{\mathrm{H}} \right]_{B=0}$. The match is a little less accurate as $\rho$ increases, or equivalently $T$ decreases. This is most likely due to higher order pole corrections which become relevant at low $T$ and one expects hydrodynamics to be less accurate. These comparisons at least prove that $\tilde{\sigma}_{\mathrm{H}}$ is nonzero in the dyonic black hole and that \eqref{Eq:Fundamentalconductivity} and \eqref{Eq:FundamentalHallconductivity} are accurate expressions for $\sigma_0$ and $\tilde{\sigma}_{\mathrm{H}}$.}

\begin{figure}[t]%
    \centering
    \subfloat[]{\includegraphics[width=0.45\textwidth]{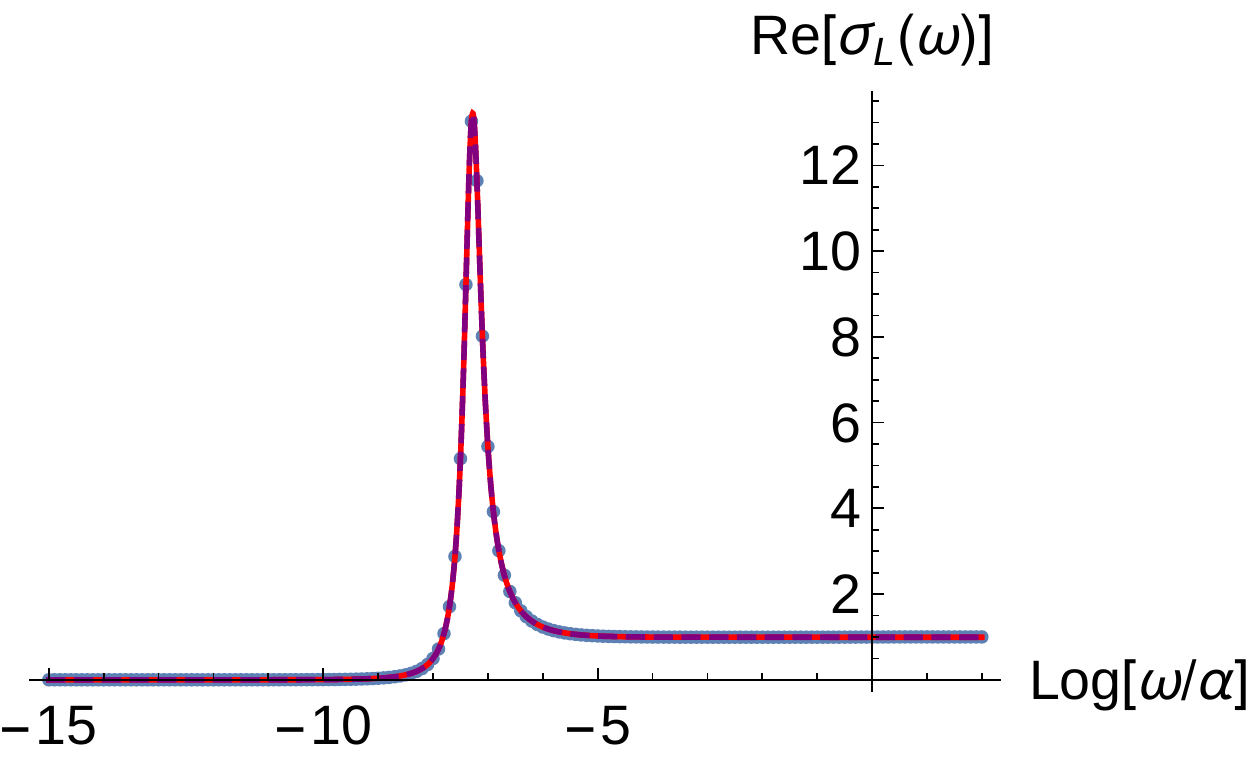}}\qquad
    \subfloat[]{\includegraphics[width=0.45\textwidth]{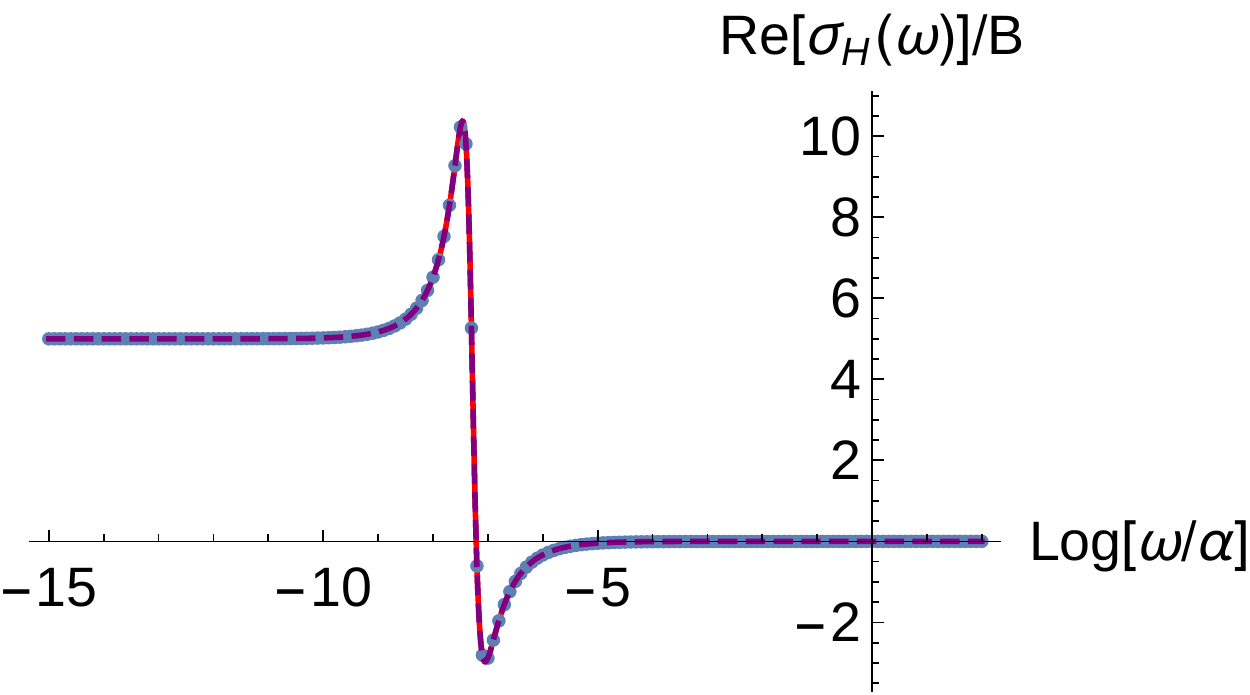}} \\
    \subfloat[]{\includegraphics[width=0.45\textwidth]{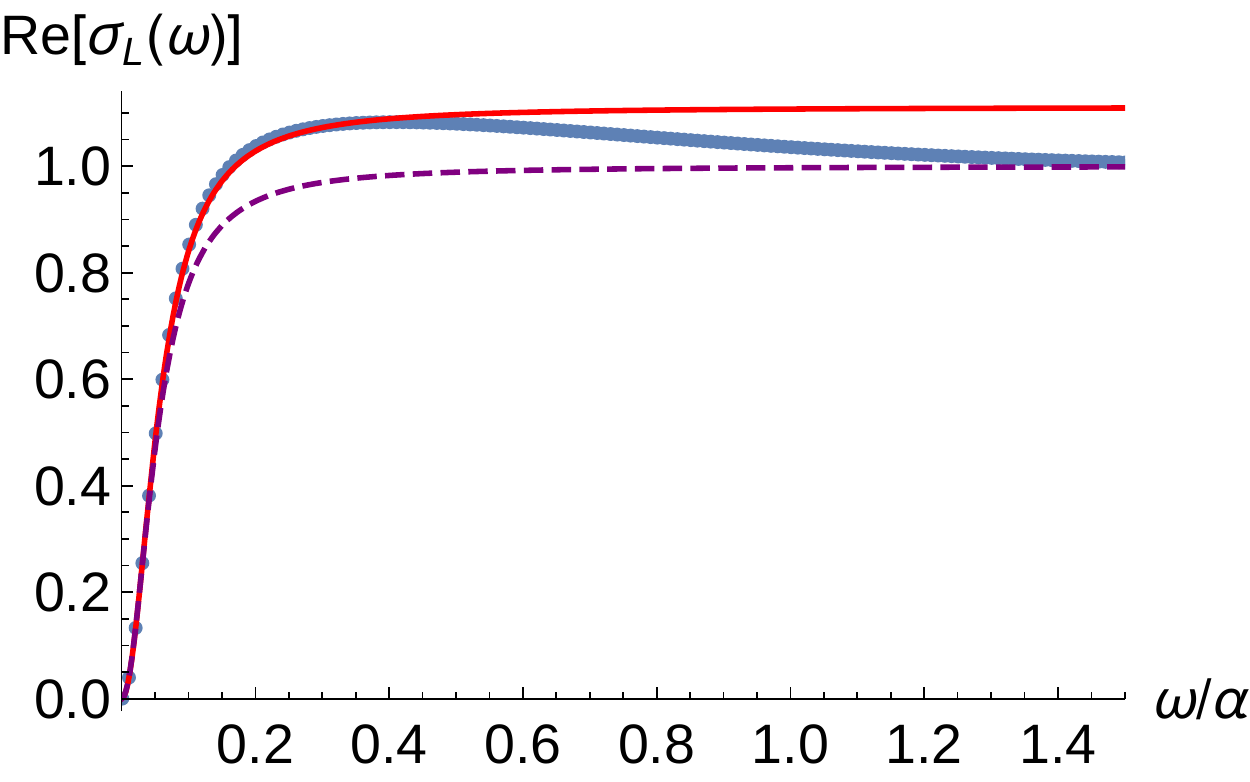}}\qquad
    \subfloat[]{\includegraphics[width=0.45\textwidth]{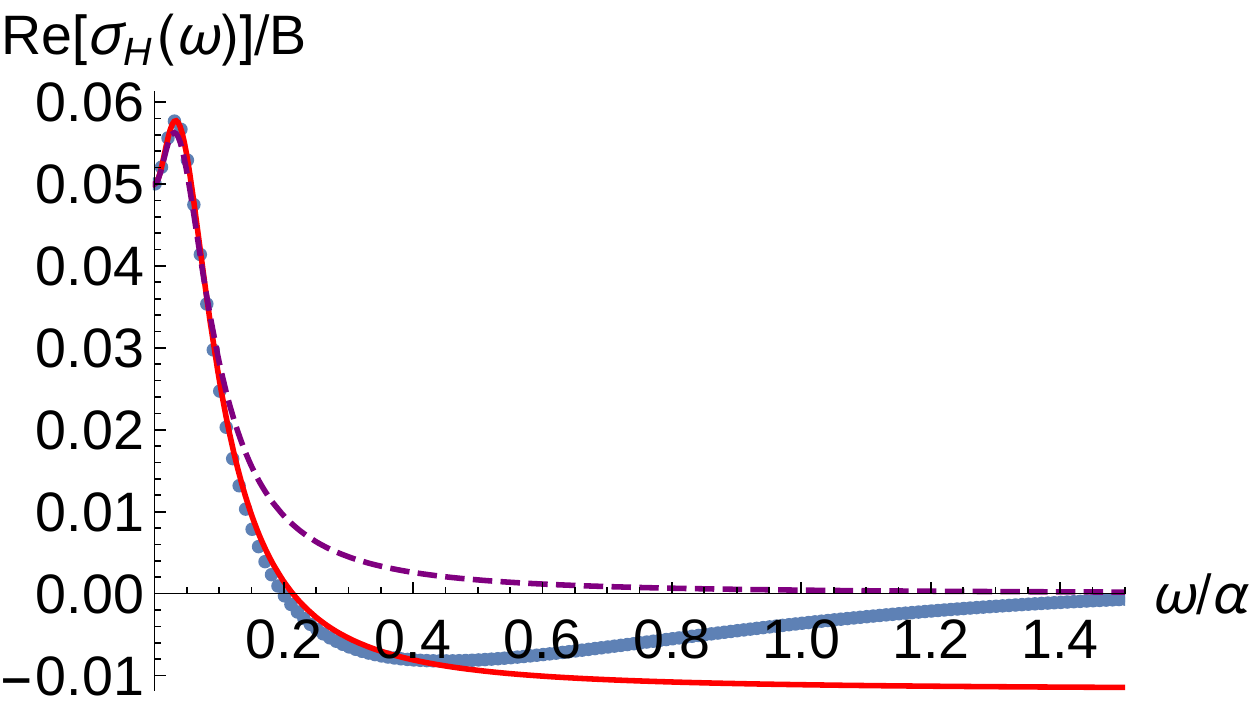}} \\
    \caption{The real parts of the AC charge conductivity as a function of frequency for two choices of charge density and magnetic field. Blue dots are data, the solid red line is our analytic result and the purple dashed line is the result of standard magnetohydrodynamics (see appendix \ref{app:standardhydro}). \textbf{Upper:} The longitudinal (left) and Hall (right) AC conductivities with $B/\alpha^2=1/100$ and $\rho/\alpha^2=1/20$. \textbf{Lower:} The longitudinal (left) and Hall (right) AC conductivities with $B/\alpha^2=1/20$ and $\rho/\alpha^2=1/100$ i.e.~$B>\rho$.}
    \label{Fig:conductivities}
\end{figure}

\subsection{Matching the correlators}
\label{sec:matching}

\begin{figure}[t]%
    \centering
    \subfloat[]{\includegraphics[width=0.45\textwidth]{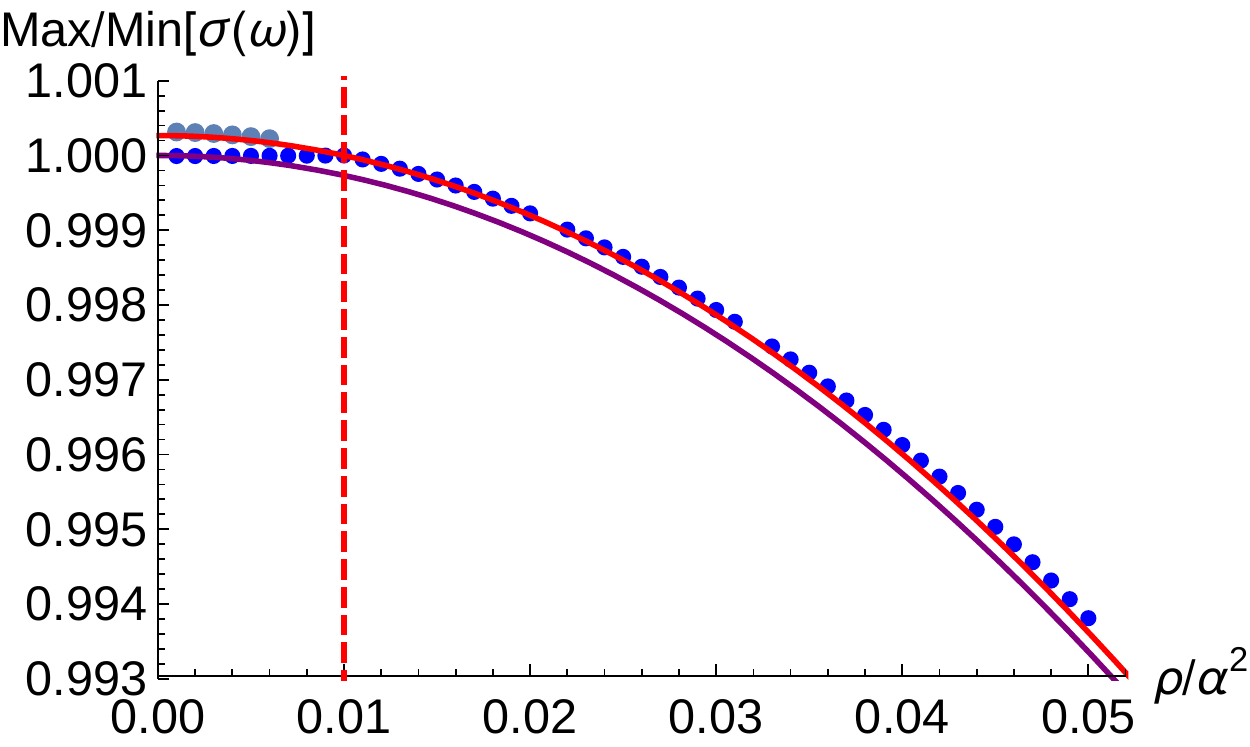}}\hfill
    \subfloat[]{\includegraphics[width=0.45\textwidth]{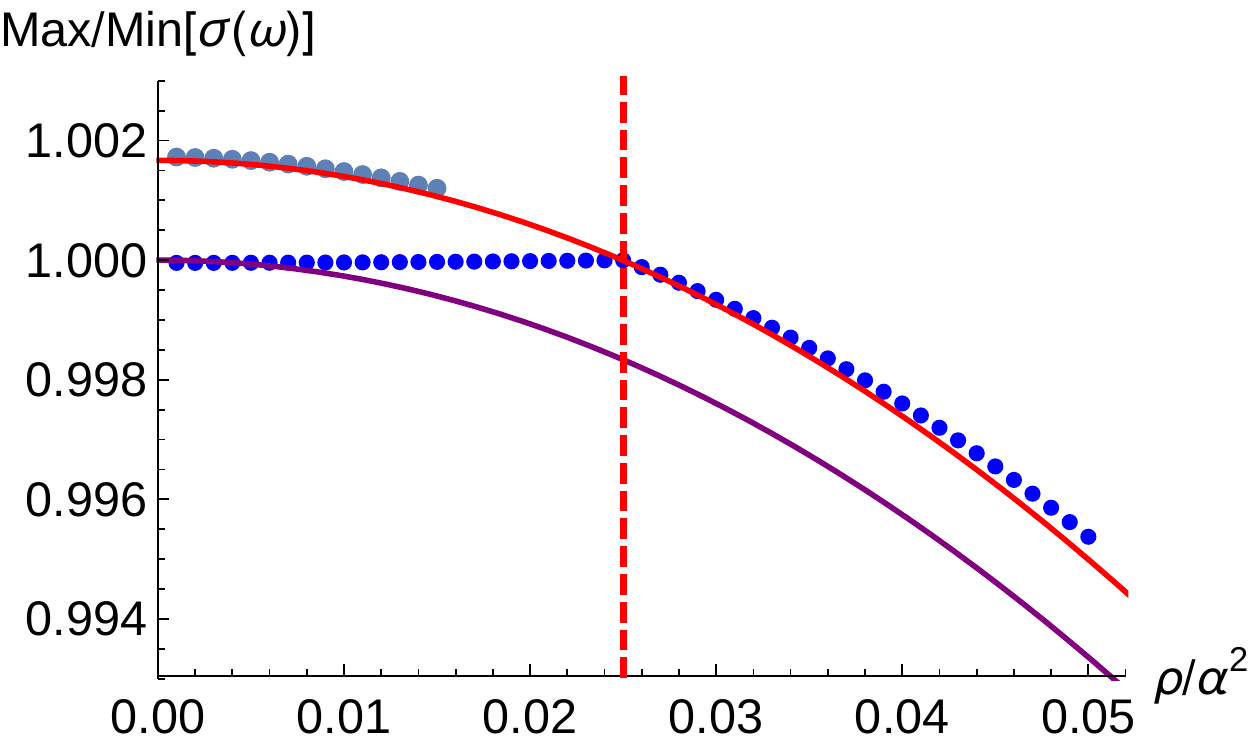}} \\
    \subfloat[]{\includegraphics[width=0.31\textwidth]{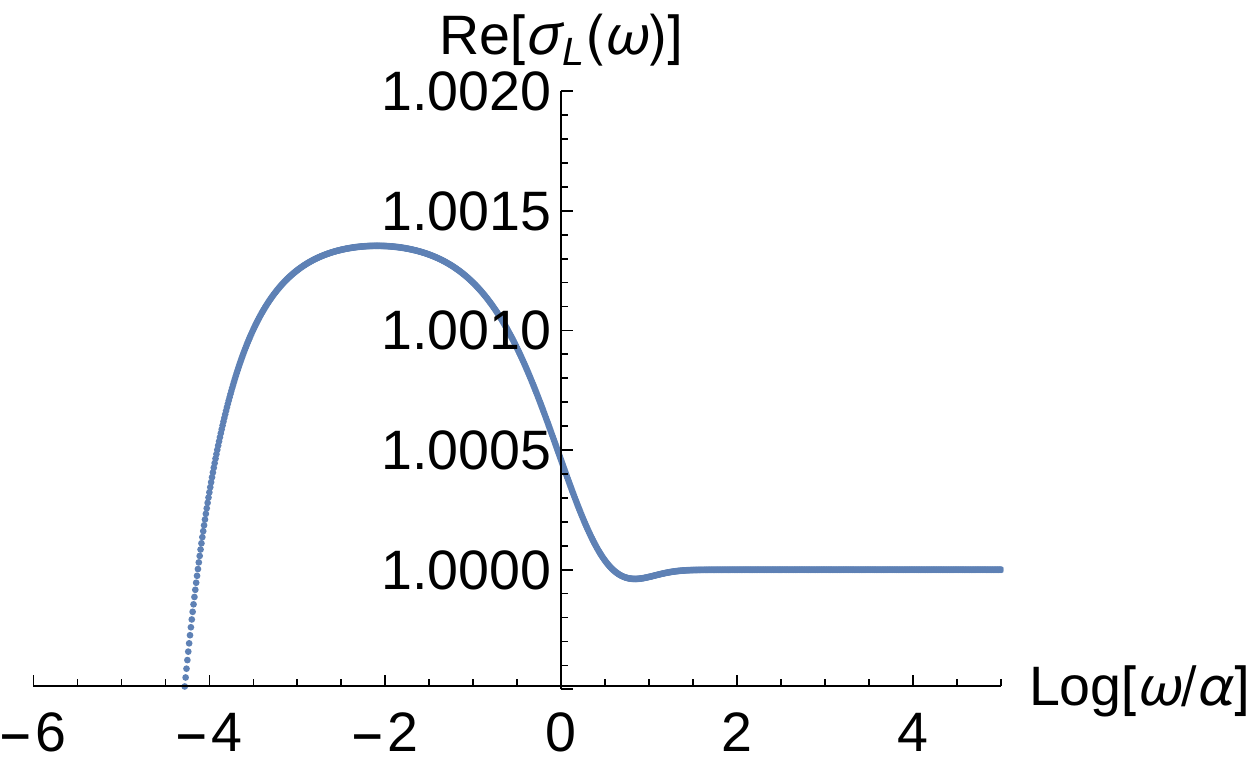}} \; 
    \subfloat[]{\includegraphics[width=0.31\textwidth]{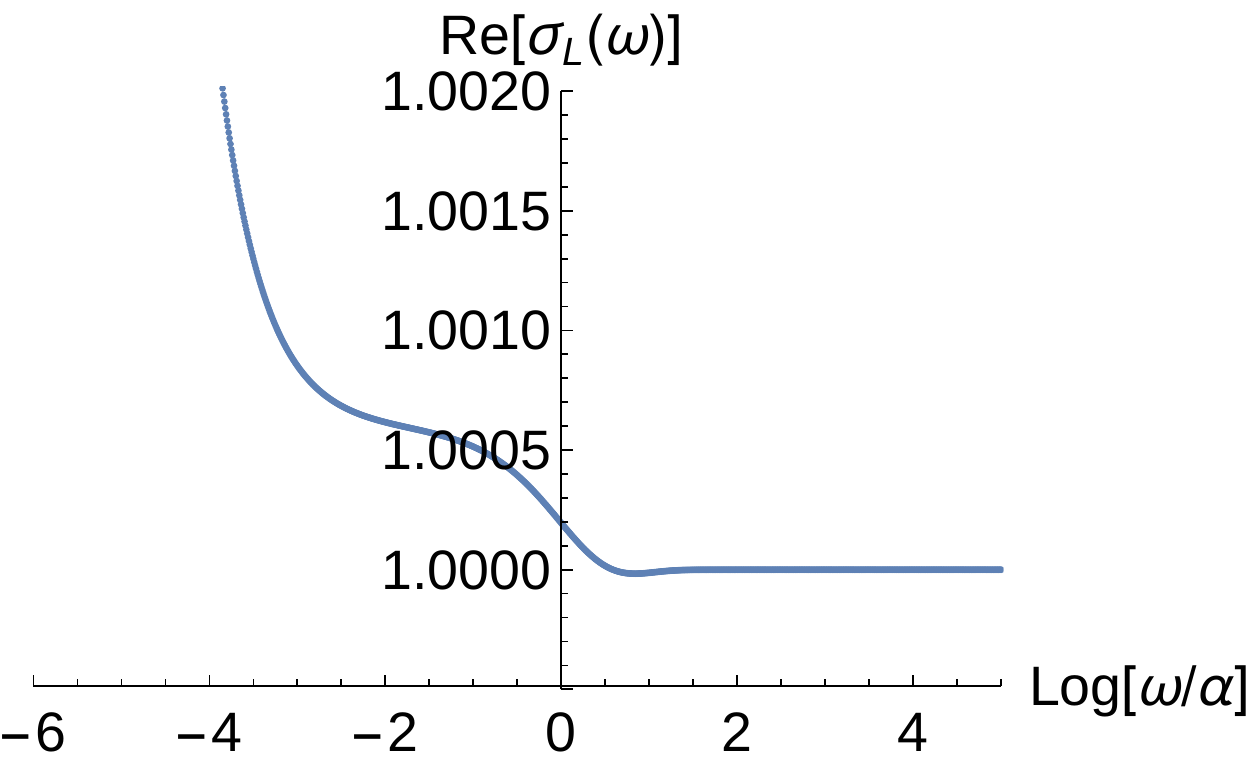}} \;
    \subfloat[]{\includegraphics[width=0.31\textwidth]{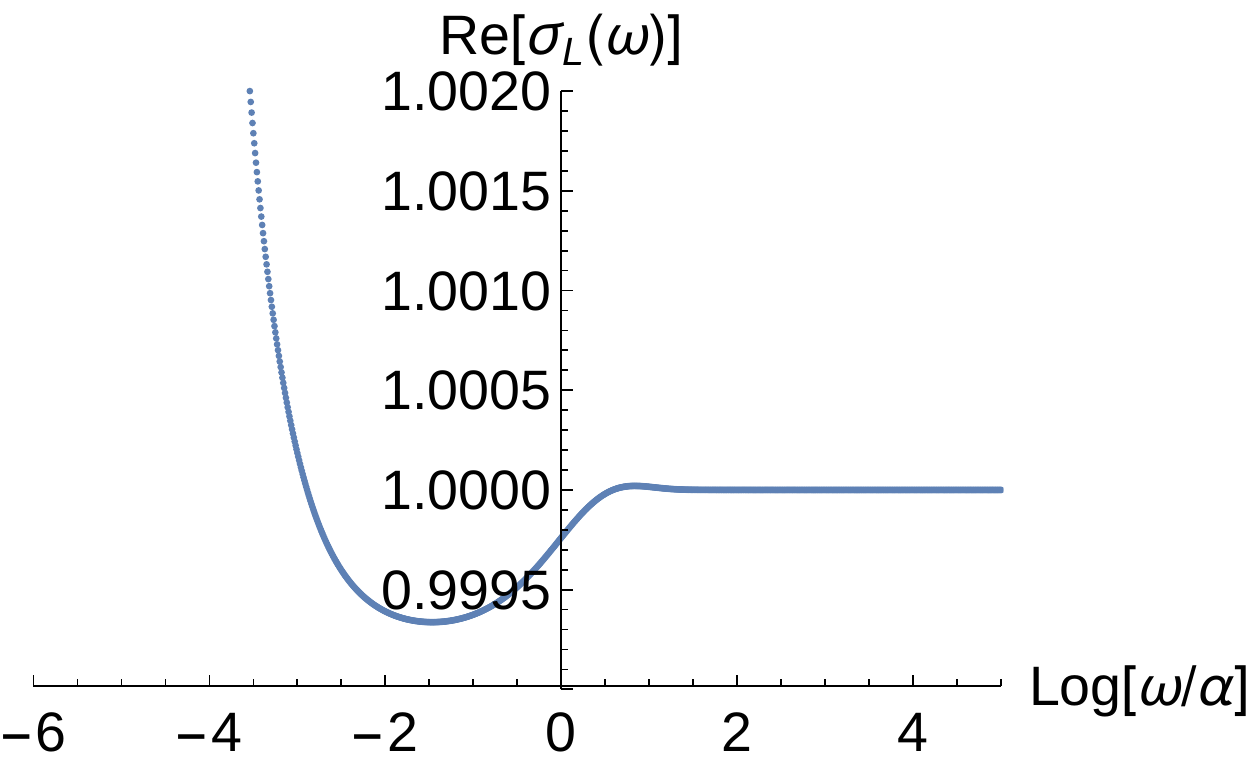}} \\
    \caption{Turning points of the AC charge conductivity as a function of charge density for two values of the magnetic field: $B/\alpha^2=0.01$ (top left) and $B/\alpha^2=0.025$ (top right). The blue dots are data while the solid red line is our analytic result for the incoherent conductivity $\sigma_{0}$. The dashed red line indicates the point where $\rho = B$ while the purple line indicates the $B=0$ limit of our analytic result, coinciding with the conductivity given by \cite{Hartnoll:2007ip}. In particular, the light blue uppermost dots of both figures which occur in the region $B > \rho$ are maxima, while the dark blue dots in the region $\rho >B$ are minima. In the bottom row we display a zoomed in plot of real part of the longitudinal charge conductivity against the logarithm of frequency at $B/\alpha^2=0.025$. The leftmost plot with $\rho/\alpha^2=1/100$ such that $B \gg \rho$ shows the local maximum at $\log \omega / \alpha \approx -2$ which corresponds to the light blue dots in the upper right figure. The rightmost plot is taken at $\rho/\alpha^2=3/100$ so that $\rho > B$ and we have a minimum at $\log \omega/\alpha \approx -1.5$. The middle plot on the bottom row indicates what happens in the intermediate region $\rho \lesssim B$.}
    \label{Fig:conductivityminimum}
\end{figure}

{\ We can now proceed to match the full correlators \eqref{Eq:SigmaL} and \eqref{Eq:SigmaH} (considering the pole position \eqref{Eq:PolePosition1s}) against the numerical results for the dyonic black hole. The outcome is shown in figure \ref{Fig:conductivities}. We distinguish two different regimes. When $\rho>B$ (figures \ref{Fig:conductivities} (a) and (b)), the agreement between \eqref{Eq:SigmaL} and \eqref{Eq:SigmaH} and the numerics is excellent in a wide range of temperature. In this case, at large $\omega$ the conductivity $\sigma_{\mathrm{L}}$ reaches a minimum before approaching the conformal value ($\sigma_{\mathrm{L}}=1$) as shown in figure \ref{Fig:conductivityminimum}. The same figure shows that the value of this minimum is very well approximated by the incoherent conductivity $\sigma_0$ defined in \eqref{Eq:Fundamentalconductivity}, which is less than 1 in this regime. The frequency at which the conductivity shows this minimum can be considered as a high frequency cut-off for the validity of the hydrodynamic regime. It is worth mentioning that, as it is evident from the purple line in figure \ref{Fig:conductivityminimum}, the $B=0$ limit of $\sigma_0$, namely the well known result $\left[ \sigma_0 \right]_{B=0}=\left[(sT/(\varepsilon + P))^2\right]_{B=0} $ of standard magnetohydrodynamics (see appendix \ref{app:standardhydro}), approximates the minimum in a significantly worse way than the full $\sigma_0$ in \eqref{Eq:FundamentalconductivitieslowB}.}

{\ In the opposite regime, $\rho <B$, the matching is good in a shorter range of frequencies as shown in figures \ref{Fig:conductivities} (c) and (d). This is reasonable since $B$ is becoming large and hydrodynamics is expected to be a worse approximation in this regime. In any case the correlators  \eqref{Eq:SigmaL} and \eqref{Eq:SigmaH} approximate the numerics consistently better than the standard magnetohydrodynamics (see appendix \ref{app:standardhydro}), which does not take into account the existence of a non-trivial $\tilde{\sigma}_{\mathrm{H}}$ (see the purple dashed line in figure \ref{Fig:conductivities} (c) and (d)). In this regime $\sigma_0>1$ as one can see in figure \ref{Fig:conductivityminimum}, and the conductivity does not show anymore a minimum at high frequency, approaching the conformal value from above. However, when $B\gg \rho$, $\sigma_0$ approximates the maximum of the conductivity, as shown in figure \ref{Fig:conductivityminimum}. Eventually, in this case the frequency at which the conductivity reaches its maximum can be defined as the UV cut-off for hydrodynamics.}

{\ We reiterate an important point in our discussion here. Hydrodynamics, like any theory, depends on a set of a priori unknown variables - the transport coefficients which must be fixed by reference to data; in our case the DC conductivities. We can of course choose different data such as the quasinormal modes to match against. However, while hydrodynamics remains a theory of a single quasinormal mode there are at most three complex constants one can fix \eqref{Eq:GenericSigma+}. Any improvement in matching against other quantities, for example the AC correlator at larger $B$ or the quasinormal mode, comes at the cost of losing an exact match with the DC conductivity.}

\section{Discussion}
\label{sec:discussion}

 In this paper we have proved that one must include a non-zero incoherent Hall conductivity $\tilde{\sigma}_{\mathrm{H}}$, in addition to the previously considered longitudinal incoherent charge conductivity $\sigma_{0}$, if one is to match hydrodynamics to the value of the DC thermal current beyond order zero in the magnetic field expansion. These incoherent conductivities - and subsequently thermo-electric correlation functions - can be expressed in terms of the DC thermal conductivities $\kappa_{\mathrm{L}}$ and $\kappa_{\mathrm{H}}$ and the thermodynamics once one appreciates that these thermal DC transport coefficients fix the $\mathcal{O}(\omega^2)$ piece of the charge current correlator. This is a consequence of the structure of the diffeomorphism and $U(1)$ gauge Ward identities \cite{Hartnoll:2007ip,Herzog_2009}. Subsequently, we have shown that this modified hydrodynamics leads to the correct effective field theory necessary to describe the hydrodynamic regime of the holographic dyonic black hole.

{\ A fundamental future direction for the present analysis will be to analyze better the role of the magnetic field $B$ in the convergence of the hydrodynamic series. In fact, in this paper we have shown that, constraining the hydrodynamic transport coefficients with the Ward identities ensures the DC limit of all the electric, thermo-electric and thermal conductivity are well described by hydrodynamics independently of the value of $B$. Moreover, the comparison between magnetohydrodynamics and the dyonic black hole performed in section \ref{sec:matching} has shown that the AC conductivities are well approximated by hydrodynamics independently of the relative value of $B$ and the charge density $\rho$. This suggests in the presence of both a temperature and a charge density, assuming that $B$ scales in the gradient expansion as a derivative might not be the correct approach. Determining the correct parameter for performing the hydrodynamic expansion, along the lines of what has been discussed at $T=0$ in \cite{Grozdanov:2016tdf}, is an issue of primary importance.}

{\ Another interesting question is to understand if the present discussion can be generalized to systems with Goldstone bosons in the presence of the magnetic field, like the charge density wave models described in \cite{Amoretti:2017axe,Amoretti:2018tzw,Amoretti:2019cef,Amoretti:2019kuf,Baggioli:2020edn}. In fact, a key assumption of the present analysis is that, as is the case in standard magnetohydrodynamics, all the correlators have a smooth $\omega\rightarrow 0$ limit. In the presence of Goldstone bosons, the correlation functions involving these fields often present poles at $\omega=0$, and analyzing how the method presented in this paper can be generalized to this case constitutes a natural question which must eventually be addressed.}

\section*{Acknowledgments}

{\ A special thanks goes to Daniel Are\'{a}n for collaboration in the early stages of this project. We thank Blaise Gout\'{e}raux for helpful comments on a previous version of this manuscript. DB would like to thank Pavel Kovtun for discussions. The project has been partially supported by the INFN Scientific Initiative SFT: “Statistical Field Theory, Low-Dimensional Systems, Integrable Models and Applications”.}

\appendix
\section{Standard formulation of relativistic magnetohydrodynamics}
\label{app:standardhydro}

{\ We take a moment here to compare our expressions with the standard versions in magneto-hydrodynamics in the Landau frame. We set $\delta u^{\mu} = (0,\vec{v})$ and note the constitutive relation for the current
    \begin{eqnarray}
        \langle J^{\mu} \rangle = q u^{\mu} + \sigma_{\mathrm{Q}} \left( F^{\mu \nu} u_{\nu} - T \Pi^{\mu \nu} \nabla_{\nu} \left( \frac{\mu}{T} \right) \right) \; .  
    \end{eqnarray}
The fluctuation of this expression around a flat background of constant $\mu$, $T$ and $B$ gives
    \begin{eqnarray}
         \delta \langle J^{\mu} \rangle 
     &=& \delta q u_{b}^{\mu} + \Pi_{b}^{\mu \nu} \left( \left( \rho \Pi_{\nu \rho}^{b} + B \sigma_{\mathrm{Q}} \Sigma_{\nu \rho}^{b} \right) \delta u^{\rho} + \delta E_{\nu} - \sigma_{\mathrm{Q}} \partial_{\nu} \delta \mu + \sigma_{\mathrm{Q}} \frac{\mu_{b}}{T_{b}}  \partial_{\nu} \delta T\right) \; . 
    \end{eqnarray}
Examining only time dependent profiles we identify the spatial part of the current
    \begin{eqnarray}
        \label{Eq:Chargecurrentfluc}
        \delta \vec{\mathcal{J}} =\left( \rho \mathbbm{1}_{2} + \sigma_{\mathrm{Q}} B \epsilon \right) \delta \vec{v} + \sigma_{\mathrm{Q}} \delta \vec{E} \; , \qquad \epsilon^2 = - \mathbbm{1}_{2} \; .
    \end{eqnarray}
From the constitutive relation of the stress-energy-momentum tensor we have
    \begin{eqnarray}
     \delta \vec{\mathcal{P}}(\omega) = (\varepsilon + P ) \delta \vec{v}(\omega) \; , 
    \end{eqnarray}
to order one in fluctuations which we can back substitute into \eqref{Eq:Chargecurrentfluc}. Employing this relationship we determine
    \begin{eqnarray}
      \label{Eq:StandardIdentificationChiSigma}
      \chi = \frac{1}{B} \left( \omega_{c} \mathbbm{1}_{2} + \gamma_{c} \epsilon \right) \; , \qquad
      \sigma_{0} = \sigma_{\mathrm{Q}} \mathbbm{1} \; , \qquad
        \omega_{c} = \frac{\rho B}{(\varepsilon + P )} \; , \qquad
        \gamma_{c} = \frac{\sigma_{0} B^2}{(\varepsilon + P )}  \; , \qquad
    \end{eqnarray}
where $\omega_{c}$ is the cyclotron frequency and $\gamma_{c}$ is the cyclotron decay rate and $\sigma_{\mathrm{Q}}=(sT/(\varepsilon+P))^2$. From these expressions it follows that
    \begin{eqnarray}
        \Gamma = \gamma_{c} \mathbbm{1}_{2} - \omega_{c} \epsilon \; , \qquad
        \Theta = \rho \mathbbm{1}_{2} + \sigma_{0} B \epsilon \; .
    \end{eqnarray}
}

{\ In the standard formulation of magnetohydrodynamics it follows from our expressions that
    \begin{eqnarray}
         \left( \Gamma - i \omega \mathbbm{1}_{2} \right)^{-1}
     &=& \frac{1}{\left( \omega + i \gamma_{c} \right)^2 -\omega_{c} ^2} \left( - \omega_{c} \epsilon  \left( i \omega -  \gamma_{c} \right) \mathbbm{1}_{2} \right) \; . 
    \end{eqnarray}
From this we determine that the charge conductivity is
    \begin{eqnarray}
         \sigma(\omega)
     &=& \sigma_{\mathrm{Q}} \frac{\omega \left( \omega + i \gamma_{c} + i \frac{\omega_{c}^2}{\gamma_{c}}\right)}{\left( \omega + i \gamma_{c} \right)^2 -\omega_{c} ^2} \mathbbm{1}_{2} 
          - \frac{\rho}{B} \frac{\omega_{c}^2 - 2 i \gamma_{c} \omega + \gamma_{c}^2}{\left( \omega + i \gamma_{c} \right)^2 -\omega_{c} ^2} \epsilon \; , 
    \end{eqnarray}
which agrees with \cite{Hartnoll:2007ip} and \cite{Hartnoll:2007ih}. We know that this expression fails to correctly evaluate the thermal conductivities except at extremely small magnetic fields.}

\section{Miscellaneous additional results}
\label{app:misc}

\begin{figure}[t]%
    \centering
    \subfloat[]{\includegraphics[width=0.45\textwidth]{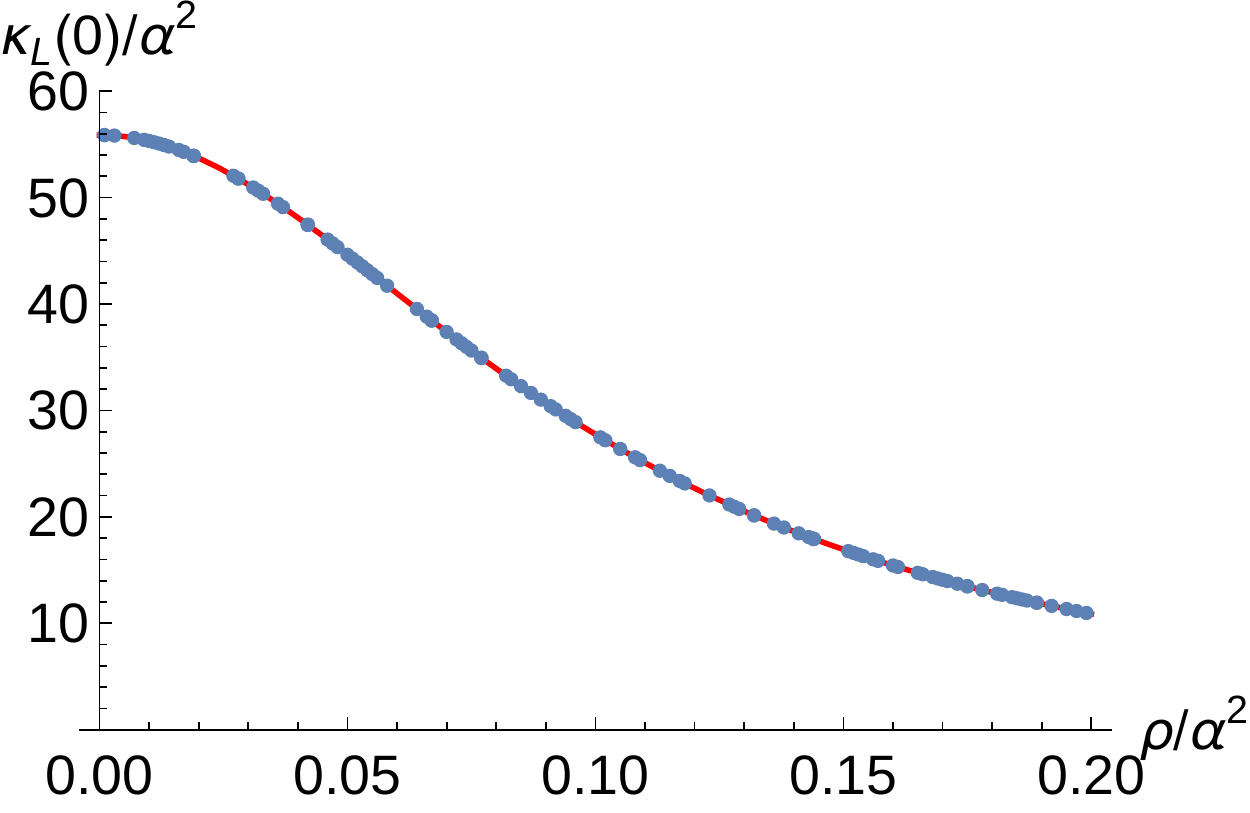}}\qquad
    \subfloat[]{\includegraphics[width=0.45\textwidth]{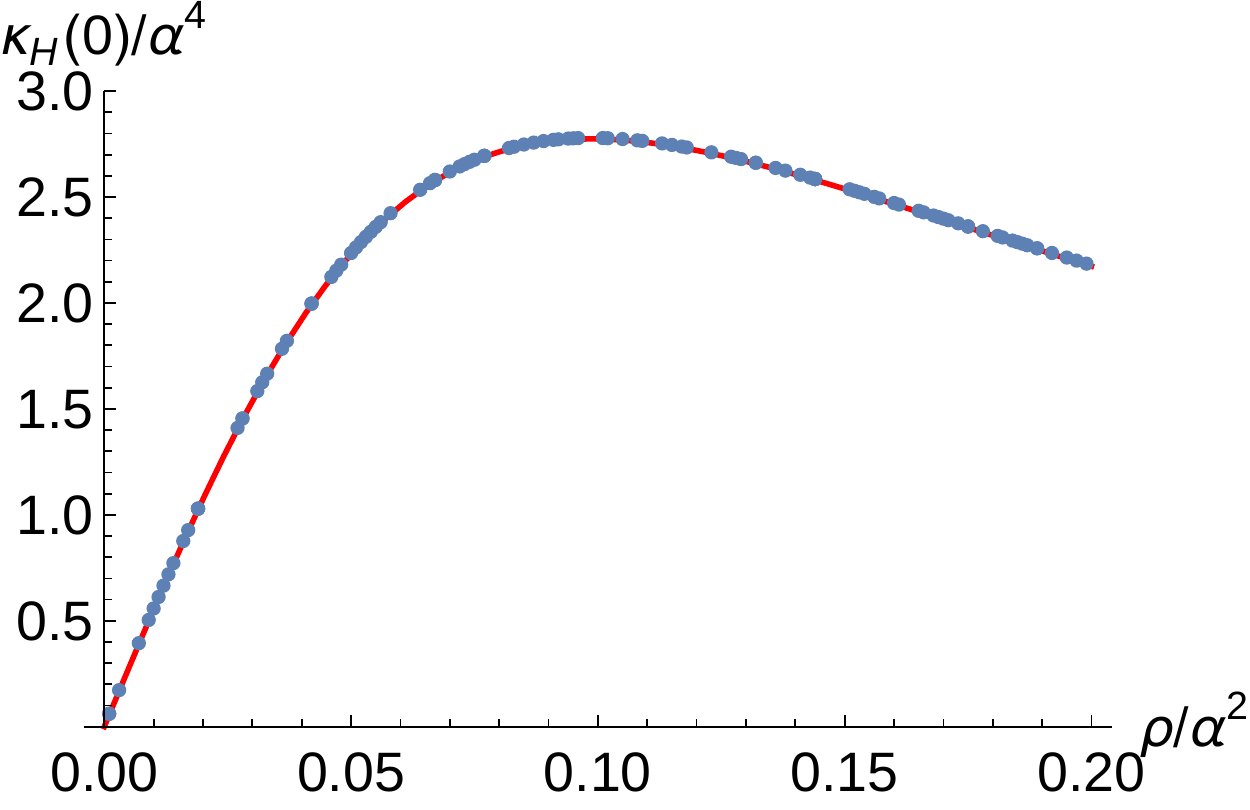}}
    \caption{Plots of the longitudinal and Hall thermal conductivites at $B/\alpha^2=1/100$ against $\rho/\alpha^2$. The blue dots represent data while the red lines are analytic expressions. The matching is generally on the order of $\sim 10^{-15}$.}
    \label{Fig:Thermalconductivities}
\end{figure}

{\ As a check on the strength of our numerics, we have extracted numerically $\kappa_{\mathrm{L}}$ and $\kappa_{\mathrm{H}}$ using the $c_2$ coefficient of the Laurent expansion around $\omega=0$,
    \begin{eqnarray}
      \label{Eq:Omega0Laurent}
      c_{2} &=& \frac{1}{2\pi} \oint_{\Gamma} d\omega \; \frac{\sigma_{+}(\omega)}{\omega^{3}}
    \end{eqnarray}
and compared to the analytical expressions \eqref{Eq:analyticalkappadyonic}. The results for  $B/\alpha^2=1/100$ as a function of $\rho/\alpha^2$ are displayed in fig.~\ref{Fig:Thermalconductivities}, showing that the analytical and numerical results match with a very high degree of accuracy. We have confirmed this for general $B$.}

\begin{figure}[t]%
    \centering
    \subfloat[]{\includegraphics[width=0.45\textwidth]{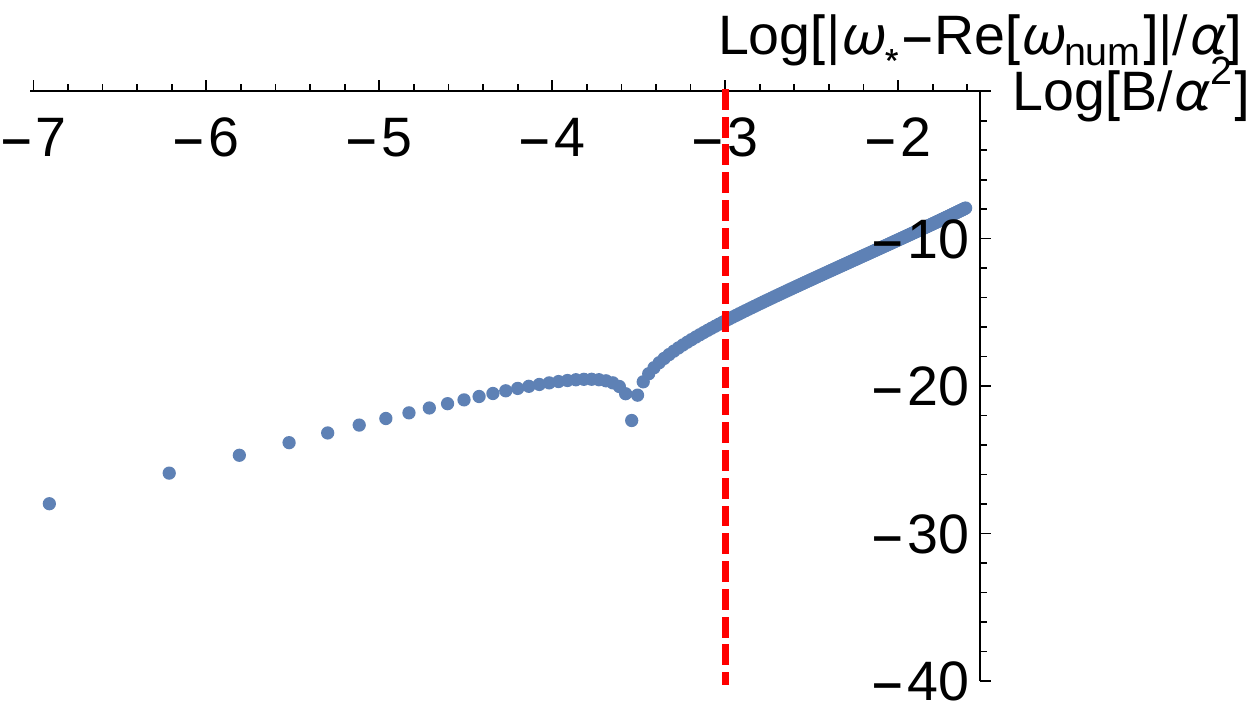}}\hfill
    \subfloat[]{\includegraphics[width=0.45\textwidth]{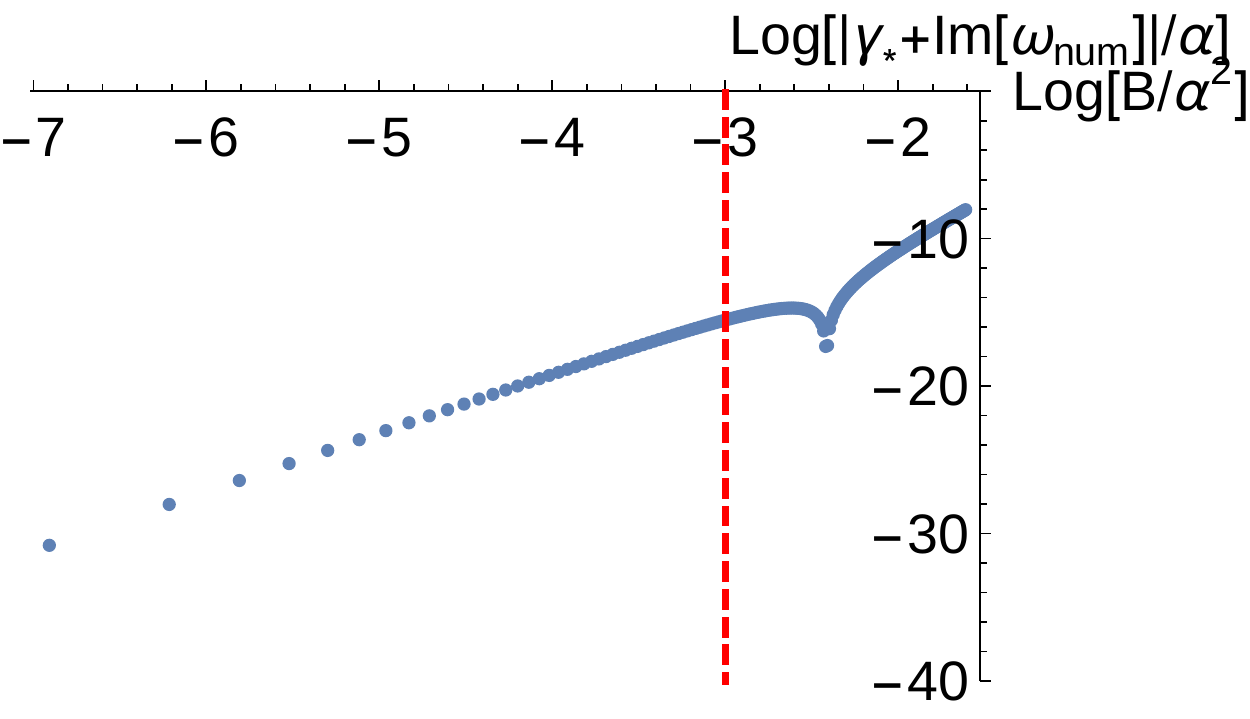}}
    \caption{Plots of the logarithm of the absolute difference between our analytic expression for the position of the hydrodynamic mode and the numerical position for the dyonic black hole against the magnetic field at $\rho/\alpha^2=1/20$. The red dashed line indicates the point where $B=\rho$. \textbf{Left:} The difference in the real part. The trough in the data indicates the point where our analytic result almost coincides with the numerical result. On the left of this trough the difference grows as $B^3$ while on the right it behaves as $B^5$. \textbf{Right:} The difference in the imaginary part. Again, the trough in the data indicates the point where our analytic result almost coincides with the numerical result. On the left of this trough the difference grows as $B^4$ while on the right it behaves as $B^6$.}
    \label{Fig:QNMs}
\end{figure}

{\ For completeness we record here the longitudinal and Hall thermo-electric and thermal AC conductivities. These are given by the expressions
    \begin{eqnarray}
     \alpha_{\mathrm{L}}(\omega) &=& \frac{i \omega (s T + \mu  \rho - m B)  \left(B \omega_{*}-\mu  \left(\gamma_{*}^2-i \gamma_{*} \omega +\omega_{*}^2\right)\right)}{B^2 \left((\omega - i \gamma_{*})^2 - \omega_{*}^2\right)} \; , \\
     \alpha_{\mathrm{H}}(\omega) &=& - \frac{ B (\omega + i \gamma_{*}) (\mu  \rho  \omega - i ( s T - m B ) \gamma_{*}) + B \omega_{*}^2 (s T - m B)}{B^2 \left((\omega - i \gamma_{*})^2 - \omega_{*}^2\right)} \nonumber \\
                                &\;& - \frac{\mu \omega_{*} (s T + \mu \rho - m B)}{B^2 \left((\omega - i \gamma_{*})^2 - \omega_{*}^2\right)} \omega ^2 \; , 
    \end{eqnarray}
and
    \begin{eqnarray}
     \kappa_{\mathrm{L}}(\omega) &=& \frac{ (s T + \mu \rho - m B) \left(B^2 (\omega + i \gamma_{*})- 2 B \mu  \omega  \omega_{*} + \mu ^2 \omega  \left(\gamma_{*}^2-i \gamma_{*} \omega +\omega_{*}^2\right)\right)}{B^2 \left((\omega - i \gamma_{*})^2 - \omega_{*}^2\right)} \; , \qquad \\
     \kappa_{\mathrm{H}}(\omega) &=& \frac{B \mu  \omega_{*}^2 (2 s T + \mu  \rho - 2 B m ) -\omega_{*} \left(B^2+\mu ^2 \omega ^2\right) (s T + \mu \rho - m B)}{B^2 \left((\omega - i \gamma_{*})^2 - \omega_{*}^2\right)} \nonumber \\
      &\;& + \frac{i \mu  ( \omega + i \gamma_{*} ) (2 B \gamma_{*} m - i \mu  \rho  ( \omega - i \gamma_{*}) -2 \gamma_{*} s T)}{B \left((\omega - i \gamma_{*})^2 - \omega_{*}^2\right)} \; ,
    \end{eqnarray}
respectively. Defining complex correlators and expanding about the pole at $\omega = \omega_{*} - i \gamma_{*}$ we find that the incoherent conductivities satisfy the relationship
    \begin{eqnarray}
      \alpha^{\mathrm{inc.}} = - \mu \sigma^{\mathrm{inc.}} \; , \qquad \kappa^{\mathrm{inc.}} = \mu^2 \sigma^{\mathrm{inc.}} \; ,
    \end{eqnarray}
which, up to the usual normalization of $\alpha^{\mathrm{inc.}}$ and $\kappa^{\mathrm{inc.}}$ by temperature (which we chose not to include in our work) is a known result. In terms of real and imaginary parts this relationship becomes
    \begin{eqnarray}
        \alpha_{0} = - \mu \sigma_{0} \; , \qquad \kappa_{0} = \mu^2 \sigma_{0} \; , \\
        \alpha_{\mathrm{H}} = - \mu \tilde{\sigma}_{\mathrm{H}} \; , \qquad \kappa_{\mathrm{H}} = \mu^2 \tilde{\sigma}_{\mathrm{H}} \; .
    \end{eqnarray}
}

{\ To get an idea of the error in our hydrodynamic charge correlator compared to the numerical charge correlator one can compare the quasi-normal mode defined by the pole in our correlator - see \eqref{Eq:PolePosition1s} and \eqref{Eq:PolePosition2s} - to the numerics. We do this in fig.~\ref{Fig:QNMs}. We can see that the accuracy and precision are quite good, although there is a systematic difference. The trough in the plots corresponds to a point where our analytic result almost matches the numerical one. To the left of this trough the analytic result overestimates the position, while to the right it underestimates.}

\bibliography{references}
\bibliographystyle{unsrt}

\end{document}